\newcommand\T{\rule{0pt}{2.6ex}}
\newcommand\B{\rule[-1.2ex]{0pt}{0pt}}
\pdfoutput=1
\documentclass[UKenglish, cernpreprint, texmf]{atlasdoc}
 
\usepackage{atlaspackage}
\usepackage{float}
\usepackage{multirow}
\usepackage{verbatim}
\usepackage{atlasbiblatex}
 
\usepackage{atlasphysics}
 
\graphicspath{{logos/}{figures/}}

\PreprintIdNumber{CERN-EP-2021-123}
\AtlasJournalRef{JHEP 12 (2021) 131}
\AtlasDOI{10.1007/JHEP12(2021)131}
 
\AtlasTitle{Measurement of $b$-quark fragmentation properties in jets using the decay $B^{\pm} \to J/\psi K^{\pm}$ in $pp$ collisions at $\sqrt{s} = 13~\TeV$ with the ATLAS detector}
 
\AtlasAbstract{
The fragmentation properties of jets containing $b$-hadrons are studied using charged $B$ mesons in 139~\ifb\ of $pp$ collisions at $\sqrt{s} = 13~\TeV$, recorded with the ATLAS detector at the LHC during the period from 2015 to 2018. The $B$ mesons are reconstructed using the decay of $B^{\pm}$ into $J/\psi K^{\pm}$, with the $J/\psi$ decaying into a pair of muons. Jets are reconstructed using the anti-$k_t$ algorithm with radius parameter $R=0.4$. The measurement determines the longitudinal and transverse momentum profiles of the reconstructed $B$ hadrons with respect to the axes of the jets to which they are geometrically associated. These distributions are measured in intervals of the jet transverse momentum, ranging from 50~$\GeV$ to above 100~$\GeV$. The results are corrected for detector effects and compared with several Monte Carlo predictions using different parton shower and hadronisation models. The results for the longitudinal and transverse profiles provide useful inputs to improve the description of heavy-flavour fragmentation in jets.
}
 
\author{The ATLAS Collaboration}
 
\AtlasRefCode{STDM-2018-52}

\hypersetup{pdftitle={ATLAS document},pdfauthor={The ATLAS Collaboration}}
 
\begin{document}
 
\maketitle
 
\tableofcontents
 
\clearpage

\section{Introduction}
\label{sec:intro}
% The next lines are included from the .//introduction.tex input file
The fragmentation of heavy quarks is a crucial aspect of quantum chromodynamics (QCD). Detailed studies and precision measurements of the heavy-quark fragmentation properties allow a deeper understanding of QCD. Furthermore, reliable modelling of the fragmentation is of great importance for measurements of the production of Higgs bosons~\cite{HIGG-2018-04,CMS-HIG-18-016}, top quarks~\cite{TOPQ-2017-12,CMS-TOP-18-002} and their associated production~\cite{HIGG-2017-03, CMS-HIG-17-026}, whose hadronic decays predominantly feature heavy quarks. Uncertainties related to the modelling of the fragmentation processes of $b$-quarks into hadrons are significant in the most precise top quark mass determinations~\cite{ATL-PHYS-PUB-2015-042, atlasTopMass7, TOPQ-2017-09, cmsTopMass8, CMS-TOP-18-004}, and are also the subject of theoretical study~\cite{corcella2002, corcella2005, corcella2010, corcella2018}. This subject has been studied in $e^+e^-$ collisions for charm~\cite{argus1991, opal1995, aleph2000, cleo2004} and bottom quarks~\cite{aleph2001, opal2003, delphi2011, sld2000, sld2002}. In hadron--hadron collisions, measurements of observables sensitive to the heavy-flavour fragmentation functions have been reported for $D^{*}$ mesons~\cite{UA1D,STARD,atlas2012}, $D^0$ mesons~\cite{aliceD0} and $J/\psi$ quarkonia~\cite{lhcbJpsi}.\\
\newline
The Monte Carlo (MC) predictions used at the LHC are tuned to describe the measurements in $e^+e^-$ collisions at relatively low centre-of-mass energies. Therefore, new measurements of $b$-quark fragmentation can be used to improve the MC simulation at LHC energy scales.\\
\newline
This analysis presents a measurement of the fragmentation of $b$-quarks into charged $B$ mesons using the ATLAS full Run~2 dataset, containing 139~\ifb\ of $pp$ collisions at $\sqrt{s}=13~\TeV$. To this end, the $B^{\pm}$ meson is reconstructed using the decay chain $B^{\pm}\to J/\psi K^{\pm} \to \mu^+ \mu^- K^{\pm}$. The procedure is similar to those in previous ATLAS measurements involving $B^{\pm}\to J/\psi K^\pm$ final states~\cite{atlasBXS, ATLAS-CONF-2015-064}. Jets are reconstructed using the anti-$k_t$ algorithm with radius parameter $R=0.4$. The reconstructed $B$ mesons are matched to jets and the longitudinal and transverse momentum profiles, $z$ and $\ptrel$, are defined as
\[
z = \frac{\vec{p}_B\cdot \vec{p}_j}{|\vec{p}_j|^2}; \ \ \ \ptrel = \frac{|\vec{p}_B\times \vec{p}_j|}{|\vec{p}_j|},
\]
where $\vec{p}_B$ is the three-momentum of the $B$ hadron and $\vec{p}_j$ is the three-momentum of the jet. The longitudinal profile, $z$, quantifies the fraction of the jet momentum carried by the $B$ meson in the direction parallel to the jet axis. On the other hand, the transverse profile, $\ptrel$, quantifies the momentum of the $B$ meson in the direction orthogonal to the jet axis. These variables are sensitive to the fragmentation function $D_q^h(x,Q^2)$, which is defined as the probability of a quark $q$ to fragment into a hadron $h$ with an energy fraction $x$ at a scale $Q$~\cite{nason,cacciari}. The measurement is performed in different intervals of the jet transverse momentum, which provides a probe of the scaling of the fragmentation functions. Furthermore, as discussed in the following, these observables are also sensitive to the contributions of gluons producing a $b\bar{b}$ pair. Since, in many cases, $b\bar{b}$ pairs arising from the splitting of high-$\pt$ gluons are not resolved into two different jets, the reconstructed $B$ meson carries a smaller fraction of the jet energy, resulting in a flatter distribution of $z$ and $\ptrel$ compared to jets with a single hard-scattering $b$-quark.\\
\newline
The paper is structured as follows. Section~\ref{sec:atlas-detector} is dedicated to the description of the ATLAS detector. A summary of the MC samples and the dataset used throughout the analysis is included in Section~\ref{sec:montecarlo}, and the object and event selection is described in detail in Section~\ref{sec:data-selection}. Section~\ref{sec:purity} is dedicated to the estimation of the purity corrections, while a discussion of the subsequent corrections for detector effects is found in Section~\ref{sec:unfolding}. The uncertainties affecting this measurement are discussed in Section~\ref{sec:systematics} and the results are presented in Section~\ref{sec:results}. Section~\ref{sec:summary} provides the summary and conclusions.
% End of text imported from the .//introduction.tex input file
\section{The ATLAS detector}
\label{sec:atlas-detector}
% The next lines are included from the .//atlas-detector.tex input file
\newcommand{\AtlasCoordFootnote}{
ATLAS uses a right-handed coordinate system with its origin at the nominal interaction point (IP) in the centre of the detector and the \(z\)-axis along the beam pipe.
The \(x\)-axis points from the IP to the centre of the LHC ring, and the \(y\)-axis points upwards. Cylindrical coordinates \((r,\phi)\) are used in the transverse plane,
\(\phi\) being the azimuthal angle around the \(z\)-axis. The pseudorapidity is defined in terms of the polar angle \(\theta\) as \(\eta = -\ln \tan(\theta/2)\). Angular distance is measured in units of \(\Delta R \equiv \sqrt{(\Delta\eta)^{2} + (\Delta\phi)^{2}}\).}
 
The ATLAS detector~\cite{PERF-2007-01} at the LHC covers nearly the entire solid angle around the collision point.\footnote{\AtlasCoordFootnote}
It consists of an inner tracking detector surrounded by a thin superconducting solenoid, electromagnetic and hadronic calorimeters,
and a muon spectrometer incorporating three large superconducting toroidal magnets with eight coils each.
 
The inner-detector system is immersed in a \SI{2}{\tesla} axial magnetic field
and provides charged-particle tracking in the range \(|\eta| < 2.5\).
The high-granularity silicon pixel detector covers the vertex region and typically provides four measurements per track,
the first hit normally being in the insertable B-layer installed before Run~2~\cite{ATLAS-TDR-2010-19,PIX-2018-001}.
It is followed by the silicon microstrip tracker, which usually provides eight measurements per track.
These silicon detectors are complemented by the transition radiation tracker (TRT),
which enables radially extended track reconstruction up to \(|\eta| = 2.0\).
The TRT also provides electron identification information
based on the fraction of hits (typically 30 in total) above a higher energy-deposit threshold corresponding to transition radiation.
 
The calorimeter system covers the pseudorapidity range \(|\eta| < 4.9\).
Within the region \(|\eta|< 3.2\), electromagnetic calorimetry is provided by barrel and
endcap high-granularity lead/liquid-argon (LAr) calorimeters,
with an additional thin LAr presampler covering \(|\eta| < 1.8\)
to correct for energy loss in material upstream of the calorimeters.
Hadronic calorimetry is provided by the steel/scintillator-tile calorimeter,
segmented into three barrel structures within \(|\eta| < 1.7\), and two copper/LAr hadronic endcap calorimeters.
The solid angle coverage is completed with forward copper/LAr and tungsten/LAr calorimeter modules
optimised for electromagnetic and hadronic measurements respectively.
 
The muon spectrometer comprises separate trigger and
high-precision tracking chambers measuring the deflection of muons in a magnetic field generated by the superconducting air-core toroids.
The field integral of the toroids ranges between \num{2.0} and \SI{6.0}{\tesla\metre}
across most of the detector.
A set of precision chambers covers the region \(|\eta| < 2.7\) with three layers of monitored drift tubes,
complemented by cathode-strip chambers in the forward region, where the background is highest.
The muon trigger system covers the range \(|\eta| < 2.4\) with resistive-plate chambers in the barrel, and thin-gap chambers in the endcap regions.\\
\newline
Interesting events are selected to be recorded by the first-level trigger system implemented in custom hardware,
followed by selections made by algorithms implemented in software in the high-level trigger~\cite{TRIG-2016-01}.
The first-level trigger accepts events from the \SI{40}{\MHz} bunch crossings at a rate below \SI{100}{\kHz},
which the high-level trigger reduces in order to record events to disk at about \SI{1}{\kHz}.
 
An extensive software suite~\cite{ATL-SOFT-PUB-2021-001} is used in the reconstruction and analysis of real and simulated data, in detector operations, and in the trigger and data acquisition systems of the experiment.
% End of text imported from the .//atlas-detector.tex input file
\section{Data and Monte Carlo samples}
\label{sec:montecarlo}
% The next lines are included from the .//montecarlo.tex input file
The dataset used in this analysis comprises the data taken from 2015 to 2018 at a centre-of-mass energy of $\sqrt{s} = 13~\TeV$. After applying quality criteria to ensure good ATLAS detector operation~\cite{DAPR-2018-01}, the total integrated luminosity used in this analysis is 139~\ifb~\cite{ATLAS-CONF-2019-021,DAPR-2013-01}. The uncertainty in the combined 2015--2018 integrated luminosity is 1.7\%~\cite{ATLAS-CONF-2019-021}, obtained using the LUCID-2 detector~\cite{LUCID2} for the primary luminosity measurements. The measurements of the fragmentation properties are unaffected by this uncertainty, given that they are normalised to the total cross section, thus cancelling out the contribution of the luminosity uncertainty. The average number of inelastic $pp$ interactions produced per bunch crossing for the dataset considered, hereafter referred to as `pile-up', is $\langle\mu\rangle = 33.6$.\\
\newline
Several different models of multijet production are used in this analysis. The MC samples were generated using the \PYTHIA[8] \cite{Sjostrand:2006za, Sjostrand:2014zea}, \SHERPA \cite{sherpa}, and \HERWIG[7] \cite{Bahr:2008pv,Bellm:2015jjp,Bellm:2017bvx} generators. These models differ in the matrix element (ME) calculation, the parton shower (PS) and the hadronisation model (HM). The main features of these samples are summarised in Table~\ref{tab:mcSamples}.\\
\newline
The \PYTHIA[8] samples were generated using \PYTHIA[8.240]. The matrix element was calculated at leading order for the $2\to 2$ process. The PS algorithm includes initial- and final-state radiation based on the dipole-style $\pt$-ordered evolution, including $g\to q\bar{q}$ branchings and a detailed treatment of the colour connections between partons \cite{Sjostrand:2006za}. The renormalisation and factorisation scales were set to the geometric mean of the squared transverse masses of the two outgoing particles (labelled 3 and 4), i.e.
\[
\muR^2 = \muF^2 = \sqrt{m^2_{\mathrm{T_3}} \cdot m^2_{\mathrm{T_4}}} = \sqrt{(\pt^2 + m_3^2) \cdot (\pt^2 + m_4^2)}.
\]
Two different sets of hadronisation and underlying-event parameter values (tunes) were used in the generation of the \PYTHIA[8] samples. Two of the samples make use of the ATLAS \textsc{A14} tune~\cite{ATL-PHYS-PUB-2014-021}, for which the \CTEQ[6L1] PDF set~\cite{cteq6l1} was used for the ME generation, the PS, and the simulation of multi-parton interactions (MPI). Two additional samples make use of the \Monash tune~\cite{monash}, interfaced to the \NNPDF[2.3lo] PDF set~\cite{nnpdf} for the ME generation, PS and MPI. The hadronisation is modelled using the Lund string model~\cite{Andersson:1983ia,Sjostrand:1984ic} in all samples. The two samples using the \textsc{A14} tune make use of the Lund--Bowler parameterisation of the fragmentation function~\cite{bowler1981}, differing in the value of the parameter $r_b$, which controls the shape of the fragmentation function for $b$-quarks: while the first of them uses the nominal value $r_b=0.855$, the second one uses the so-called \textsc{A14-rb} tune, for which $r_b=1.05$, as obtained from a combined fit to LEP and SLD data~\cite{aleph2001,opal2003,delphi2011,sld2000}. The two samples using the \Monash tune make use of the Peterson~\cite{peterson1983} and Lund--Bowler~\cite{bowler1981} parameterisations for the fragmentation functions, where the $r_b$ parameter is set to $r_b = 0.855$ for the Lund--Bowler sample. The \PYTHIA[8] sample making use of the \textsc{A14-rb} tune is taken as the nominal MC sample, used across the analysis for unfolding and uncertainty estimation.\\
\newline
The \SHERPA samples were generated using \SHERPA[2.2.5]. The matrix element was calculated at leading order (LO) for the $2\to 2$ process, and the default \SHERPA CSS dipole PS~\cite{Catani:2001cc, Krauss:2002up}. Matrix element renormalisation and factorisation scales for $2\to 2$ processes were set to the harmonic mean of the Mandelstam variables $s$, $t$ and $u$~\cite{mandelstam}. The \CT[14nnlo]~\cite{ct14} PDF set was used for the matrix element calculation, as well as for the modelling of the PS and MPI. The different \SHERPA samples use different hadronisation models. The first one makes use of the \SHERPA AHADIC model for hadronisation \cite{Winter:2003tt}, which is based on the cluster hadronisation algorithm~\cite{Webber:1983if}, while the second one uses the Lund string model~\cite{Sjostrand:2006za, Bothmann:2019yzt,Andersson:1983ia,Sjostrand:1984ic} for the modelling of the hadronisation.\\
\newline
Finally, two \HERWIG[7] samples were generated at leading order using \HERWIG[7.2.1]. The matrix element for the $2\to 2$ process was calculated at LO with the \MMHT[lo] PDF~\cite{mmht}. The renormalisation and factorisation scales were set as
\[
\muR^2 = \muF^2 = \frac{2stu}{s^2+t^2+u^2},
\]
where $s$, $t$ and $u$ are the Mandelstam variables~\cite{mandelstam}. The first sample uses the default \HERWIG[7] angle-ordered PS, while the second sample uses a dipole-based PS~\cite{Platzer:2011bc}. For both \HERWIG[7] samples, the \MMHT[lo]~\cite{mmht} PDF was used for the modelling of the MPI, and the hadronisation was modelled by means of the default \HERWIG[7] cluster hadronisation algorithm.\\
\newline
The decays of the $B$ mesons, of particular importance for this analysis, were modelled using the \EVTGEN[1.6.0] generator~\cite{evtgen} for the totality of the samples described above. The samples simulated using \PYTHIA[8]~\textsc{A14-rb}, \PYTHIA[8]~\textsc{A14}, and \SHERPA with string hadronisation were passed through the $\GEANT$-based~\cite{Agostinelli:2002hh} ATLAS detector-simulation program~\cite{SOFT-2010-01} since they were also used to correct the measurements for detector effects, as described in Section~\ref{sec:unfolding}. They are reconstructed and analysed with the same processing chain as the data. These fully simulated samples include the effect of multiple $pp$ interactions per bunch crossing, simulated using \PYTHIA[8.186] interfaced to the \textsc{A3} tune~\cite{ATL-PHYS-PUB-2016-017}, as well as the effect on the detector response of interactions from bunch crossings before or after the one containing the hard interaction. In addition, during the data-taking, some modules of the ATLAS hadronic calorimeter were disabled for some periods of time. The resulting non-functioning regions in the hadronic calorimeter were not necessarily included in the simulation for all the samples. However, the effect of removing jets pointing towards these regions has been found to be negligible.
\begin{table}[h]
\small
\caption{Properties of the Monte Carlo samples used in the analysis, including the perturbative order in \alphas, the number of final-state partons, the PDF set, the parton shower algorithm, the renormalisation and factorisation scales, the tune and the hadronisation model. Further details can be found in the text.}
\begin{center}
\begin{tabular}{c|c|c|c|c|c|c}
\T\B Generator & ME order & Scales \muR, \muF & Parton shower & PDF set & Tune & Hadronisation\\
\hline
\hline
\multirow{4}{*}{\T\T\B \PYTHIA[8]} & \multirow{4}{*}{$2\to 2$ @ LO} & \multirow{4}{*}{$(m_{\mathrm{T}3}\cdot m_{\mathrm{T}4})^{\frac{1}{2}}$} & \multirow{4}{*}{$\pt$-ordered} & \multirow{2}{*}{\CTEQ[6L1]}   & \T \textsc{A14} & Lund--Bowler\\
&                              &                             & &                            & \B \textsc{A14-rb} & Lund--Bowler\\
&                              &                             & & \multirow{2}{*}{\NNPDF[2.3]} & \multirow{2}{*}{\Monash} & \T Lund--Bowler\\
&                              &                             & &                            &                         & \B Peterson\\
\hline
\multirow{2}{*}{\T\B \SHERPA} & \multirow{2}{*}{$2\to 2$ @ LO} & \multirow{2}{*}{$H(s,t,u)$} & \multirow{2}{*}{CSS (dipole)}  & \multirow{2}{*}{\CT[14]} & \multirow{2}{*}{--} & \T\B Cluster model\\
&                              &                             & &                             &                        & \B Lund string model\\
\hline
\multirow{2}{*}{\T\B \HERWIG[7]} & \multirow{2}{*}{$2\to 2$ @ LO} & \multirow{2}{*}{$\sqrt{\frac{2stu}{s^2+t^2+u^2}}$} & \T Angle-ordered & \multirow{2}{*}{\MMHT} & \multirow{2}{*}{--} & \multirow{2}{*}{Cluster model}\\
 
&                              &                             & \B Dipole & & &\\
\end{tabular}
\end{center}
\label{tab:mcSamples}
\end{table}

% End of text imported from the .//montecarlo.tex input file
\section{Object and event selection}
\label{sec:data-selection}
% The next lines are included from the .//data-selection.tex input file
Events are selected using triggers optimised for $J/\psi$ meson identification and selection in its decay into muon pairs~\cite{TRIG-2018-01}. These triggers select events with two muons with $\pt > 6~\GeV$. During the 2017 and 2018 data-taking periods, the trigger also required the invariant mass of the dimuon pair to satisfy $2~\GeV < m_{\mu\mu} < 9~\GeV$ and the angular separation between the muons to satisfy $\Delta R_{\mu\mu} < 1.5$. The trigger selection efficiency is about 60\%.\\
\newline
The primary vertex of the event is reconstructed as the vertex maximising the value of $\sum\pt^2$ for all tracks originating from it. The reconstruction of $B^\pm$ mesons is done by re-fitting a pair of oppositely charged muons with $\pt > 6~\GeV$ and $|\eta| < 2.5$ to a common vertex, with a fit quality fulfilling $\chi^2 < 10$. Both muons are reconstructed using information from both the inner detector and the muon spectrometer. They must pass the \textsc{Medium} quality requirements~\cite{muons, MUON-2018-03}, including at least one hit in the precision chambers of the muon spectrometer, and are required to have an invariant mass consistent with the mass of the $J/\psi$ meson, i.e.\ to lie within the range 2.6--3.6~\GeV. Moreover, both muons must spatially match the muon objects used in the trigger selection within $\Delta R = 0.01$. Given the short $J/\psi$ lifetime following the $B^{\pm}\to J/\psi K^\pm$ decay, a third track with $\pt> 4~\GeV$ and $|\eta| < 2.5$ is fitted to a common vertex together with the two muon tracks, requiring $\chi^2 / N_{\mathrm{dof}} < 2.0$. In the fit, the invariant mass of the pair of muon tracks is constrained~\cite{ATL-PHYS-2003-031} to match the world average value for the $J/\psi$ mass, $m_{J/\psi} = 3096.900~\MeV$~\cite{pdg2018}. Moreover, each of the three tracks in the triplet must have at least one hit in the pixel detector and at least four hits in the silicon microstrip detector, while the invariant mass of the triplet is required to lie within the range 5.0--5.7~\GeV. Finally, the pseudo-proper lifetime of the $B^\pm$ candidate, $\tau = m_B L_{xy} / \pt$, where $m_B$ is the PDG value of the $B^\pm$ mass, $m_B = 5279.320~\MeV$~\cite{pdg2018}, and $L_{xy}$ is the transverse flight distance from the primary vertex, is required to be above 0.20 ps. The efficiency of this cut has been shown to be well described by the MC simulations, and hence to have only a small impact on the results after the corrections for detector effects are applied. The mass of the charged kaon, $m_K = 493.677~\MeV$~\cite{pdg2018}, is assumed for the third track for the sake of calculating the three-body invariant mass.\\
\newline
Jet reconstruction is performed using particle-flow objects~\cite{pflow,JETM-2018-05} as inputs to the anti-$k_t$ algorithm as implemented in FastJet~\cite{antikt,fastjet} with a radius parameter $R=0.4$. Muon tracks are not considered in the particle-flow algorithm and, thus, the energy of muons from the $B^\pm$ decay chain is not accounted for in the jet energy. In order to correct for the presence of muons inside the jets, muons passing the same quality requirements as those used in the $B^\pm$ reconstruction are matched to the jet by means of the ghost-association method~\cite{jetAreas}. This procedure defines objects with infinitesimal energy and the same direction as the track momentum (ghosts), which are then used as inputs to the jet reconstruction algorithm. A track is considered matched to a given jet if the ghost associated with it is clustered inside the jet. The jet four-momentum is then corrected by vectorially adding to it the four momenta of the muons associated with the jet, after subtracting the muon energy loss in the calorimeter. After this correction is applied, the jet energy is calibrated and, in order to ensure full inner-detector acceptance, jets with $\pt > 20~\GeV$ and $|\eta| < 2.1$ are preselected. The jet calibration procedure includes energy corrections for pile-up, as well as angular corrections. Effects due to energy losses in inactive material, shower leakage, the magnetic field and inefficiencies in energy clustering and jet reconstruction are taken into account. This is done using a simulation-based correction, in bins of $\eta$ and $\pt$, derived from the relation of the reconstructed jet energy to the energy of the corresponding particle-level jet. In a final step, an in situ calibration corrects for residual differences in the jet response between the MC simulation and the data using $\pt$-balance techniques for dijet, $\gamma$+jet, $Z$+jet and multijet final states. In order to reject pile-up jets, the so-called `jet vertex tagger' (JVT) algorithm is used~\cite{jvt}. Moreover, in order to avoid overlaps between jets (and thus, incorrect matchings with $B$ mesons), jets at a distance $\Delta R < 0.8$ from any other jet with $\pt > 20~\GeV$ are discarded.\\
\newline
The reconstructed $B^\pm$ mesons are then matched to the corresponding jet by requiring both objects to be within $\Delta R = 0.4$ of each other. If there is more than one $B$ meson candidate for a single jet, the one with lowest $\chi^2$ to the three-track vertex is selected. If any of the three tracks arising from the $B^\pm$ decay is found to lie at $\Delta R > 0.4$ from the jet, its four-momentum is added to the jet four-momentum. The basic object for this analysis is, therefore, a jet containing a $B^\pm$ meson. The selection criteria described above are met by 1\,413\,684 jets in 1\,404\,620 events.\\
\newline
To correct the simulation for differences in the muon identification and JVT acceptance efficiencies from those measured in the data, simulated events are weighted using dedicated corrections provided to that end \cite{muons, jvt}. The differences in the trigger efficiencies between data and the simulation are accounted for as a systematic uncertainty, as described in Section~\ref{sec:systematics}.\\
\newline
Figure~\ref{fig:detLevel} shows the detector-level distributions of the jet $\pt$ and the $\pt$ of the $B^\pm$ meson within the jet, together with the MC predictions. The data distributions include the purity corrections described in Section~\ref{sec:purity}. While the \SHERPA sample with string-based fragmentation describes the jet $\pt$ within 10\%, it fails to describe the $B$ meson $\pt$, with discrepancies of over 50\%. The \PYTHIA samples with the \textsc{A14} and \textsc{A14-rb} tunes give similar descriptions of the jet $\pt$, with differences from data of up to 40\% in the tail of the distribution. The differences between the two tunes are clearly observed in the distribution of the $B$ meson $\pt$, where \textsc{A14} gives a more accurate description within 10\% while \textsc{A14-rb} shows discrepancies of up to 30\% in the tail of the distribution.
\begin{figure}[H]
\includegraphics[width=0.50\textwidth]{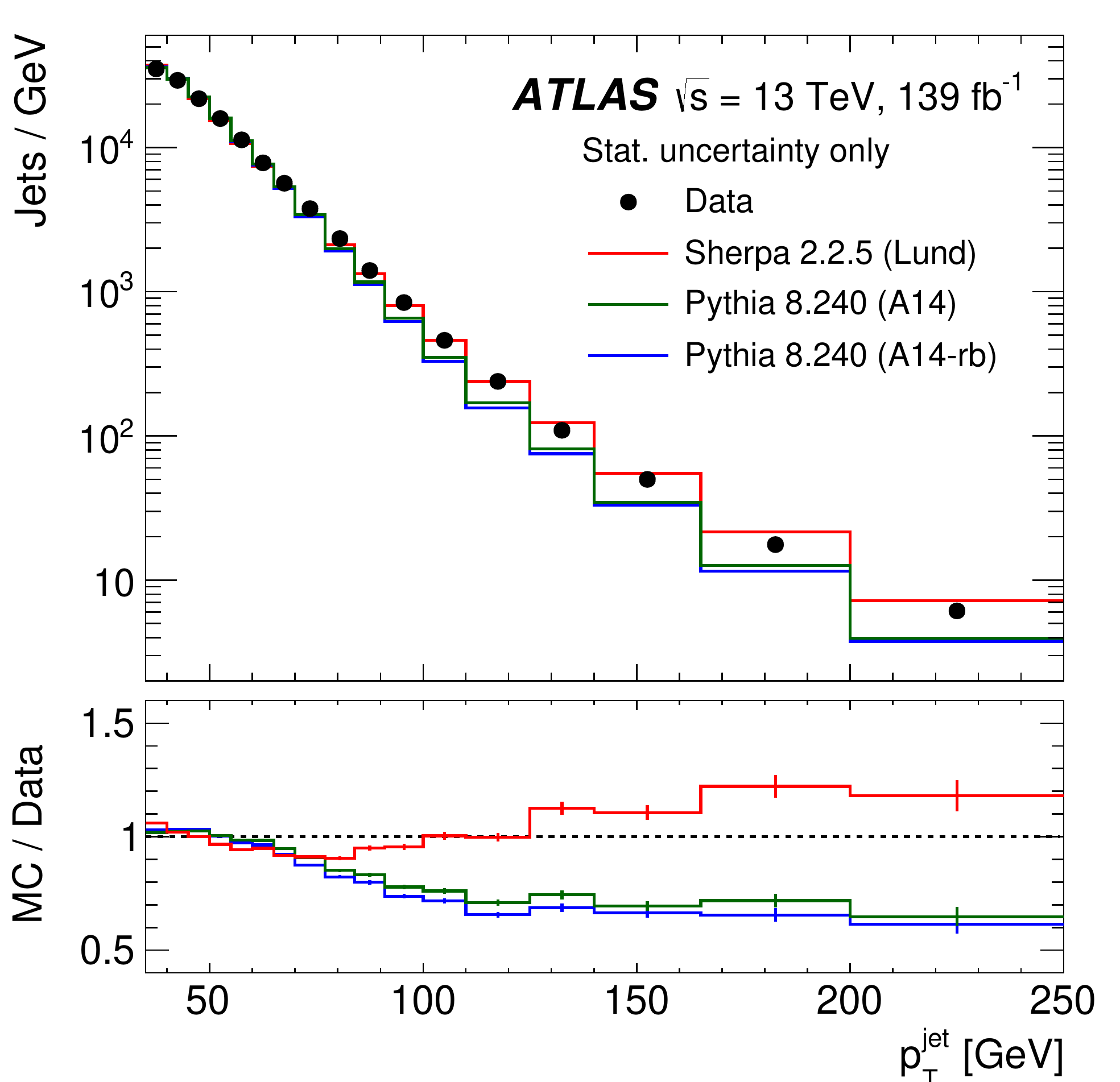}
\includegraphics[width=0.50\textwidth]{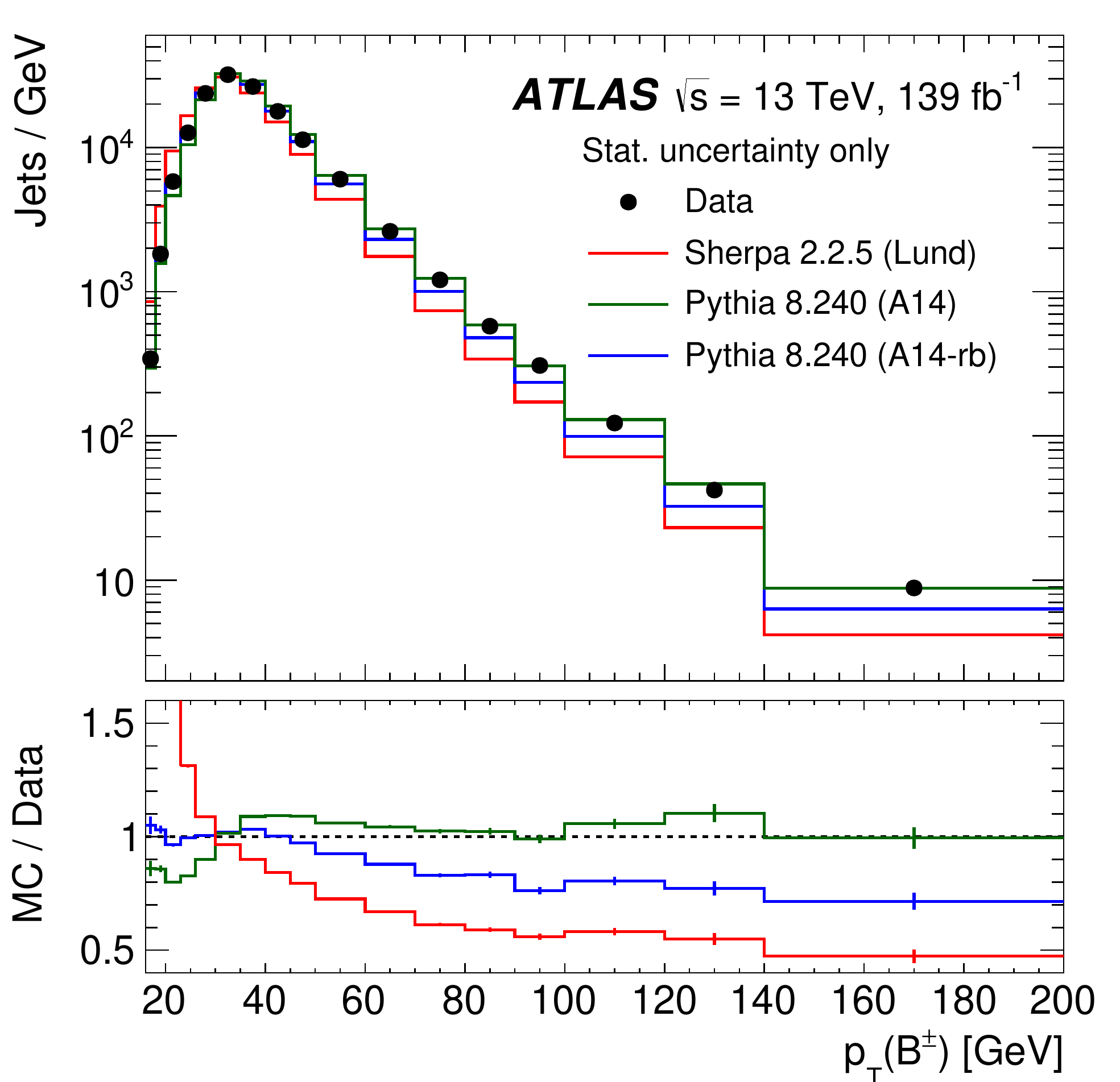}
\caption{Distributions of the $\pt$ of the selected jets and the $B^\pm$ meson candidates within them, corrected for purity effects, together with MC predictions. The vertical error bars show the statistical uncertainties only.}
\label{fig:detLevel}
\end{figure}
 
For the fragmentation measurement, the data are binned in three intervals of the jet $\pt$, namely $[50, 70)$, $[70, 100)$ and $\pt \geq 100~\GeV$. This choice ensures that the bin width is larger than the resolution in the different $\pt$ intervals~\cite{pflow}. For each of these intervals, the $z$ distributions are binned in intervals with a width of 0.07, while the $\ptrel$ distributions use a variable bin width. This choice allows finer binning, while retaining sufficient statistical precision in each bin, as needed for the estimation of the signal purity described in Section~\ref{sec:purity}. The lower limit of the $z$ distribution is extended towards lower values with increasing jet $\pt$, where a larger fiducial phase space becomes available. The $z$ and $\ptrel$ distributions are normalised to unit area by dividing by the number of entries, allowing the systematic uncertainties to be reduced.\\
\newline
Since the momentum resolution for charged particles is much better than the jet energy resolution, the measured $z$ distribution extends beyond unity. The values larger than one are absorbed as an overflow into the last bin of the $z$ distribution. Similarly, values of $z$ below the lower limit of the first bin are absorbed into the first bin as an underflow, and values of $\ptrel$ above the upper limit of the last bin are absorbed into it as an overflow.
% End of text imported from the .//data-selection.tex input file
\section{Signal extraction}
\label{sec:purity}
% The next lines are included from the .//purity.tex input file
The selected sample of $B^\pm$ meson candidates does not only contain $B^\pm$ mesons, but also backgrounds arising from different sources. For each $(\pt,x)$ bin, with $\pt$ being the jet $\pt$ and $x=z$ or $\ptrel$, only a fraction $p(\pt, x)$ of the reconstructed entries, referred to as the purity, are real $B^\pm$ mesons. The number of $B^{\pm}$ mesons, $R$, reconstructed in a given bin can thus be expressed as
\[
R(\pt, x) = N(\pt,x)\times p(\pt, x),
\]
where $N$ is the total number of candidates in a given bin. In order to determine $p(\pt,x)$, a binned maximum-likelihood fit to the invariant mass distribution of the $B^\pm$ candidates is performed. The probability density function (pdf) of the model describing the invariant mass distribution can be written as a combination of the signal and background pdfs as
\[
\mathcal{F}(m) = \lambda_s \mathcal{F}_s(m) + \lambda_{B_x} \mathcal{F}_{B_x}(m) + \lambda_{B_\pi} \mathcal{F}_{B_\pi}(m) + \lambda_c\mathcal{F}_{c}(m),
\]
where $\mathcal{F}_s$, $\mathcal{F}_{B_x}$, $\mathcal{F}_{B_\pi}$ and $\mathcal{F}_{c}$ are the pdfs for each of the components, signal or background, and $\lambda_s$, $\lambda_{B_x}$, $\lambda_{B_\pi}$ and $\lambda_{c}$ are coefficients representing their relative fractions, and thus $\lambda_s(\pt,x) = p(\pt,x)$. The closure relation $\sum_i \lambda_i = 1$ must be satisfied by these coefficients. Each of the fit components is described below.
\begin{itemize}
\item{} The signal $\mathcal{F}_s$, arising from the real $B^\pm$ meson contribution, is modelled by a double-Gaussian function
\[
\mathcal{F}_{s}(m | \mu,\sigma_1,\sigma_2, \beta) = \frac{1}{\sqrt{2\pi}}\left\{\frac{\beta}{\sigma_1} \exp\left[-\frac{(m-\mu)^2}{2\sigma_1^2}\right] + \frac{1-\beta}{\sigma_2} \exp\left[-\frac{(m-\mu)^2}{2\sigma_2^2}\right]\right\},
\]
where $\beta$ is the relative normalisation of the two Gaussian components.
\item{} The misreconstructed background, $\mathcal{F}_{B_x}$, arising from the decays $B^{\pm/0} \to J/\psi K^{*\pm/0} \to J/\psi (K\pi)^{\pm/0}$ and $B^{\pm/0}\to J/\psi (K\pi)^{\pm/0}$, creates a low-mass structure, displaced from the $B^\pm$ mass. It is modelled by the following function
\[
\mathcal{F}_{B_x}(m | b,s) = 1-\tanh\left(\frac{m-s}{b}\right).
\]
\item{} The resonant background $\mathcal{F}_{B_\pi}$, arising from the decays $B^\pm\to J/\psi \pi^\pm$, creates a peaking structure displaced from the $B^\pm$ mass towards higher masses. It is modelled by the sum of a Gaussian and an asymmetric Gaussian function,
\begin{equation}
\mathcal{F}_{B_\pi}(m | \mu_1, \mu_2, \hat\sigma_1, \hat\sigma_2, \hat\sigma_3, \gamma) = \frac{1}{\sqrt{2\pi}}\left\{\frac{\gamma}{\hat\sigma_1} \exp\left[-\frac{(m-\mu_1)^2}{2\hat\sigma_1^2}\right] + (1-\gamma)\mathcal{G}_\mathrm{asym}(m| \mu_2,\hat\sigma_2,\hat\sigma_3)\right\},
\label{eq:pionBkg}
\end{equation}
where $\gamma$ is a relative normalisation factor and
\[
\mathcal{G}_\mathrm{asym}(m | \mu_2,\hat\sigma_2,\hat\sigma_3) = \left\{
\begin{array}{l}
\displaystyle{\frac{1}{\hat\sigma_2}\exp\left[-\frac{(m-\mu_2)^2}{2\hat\sigma_2^2}\right] \mbox{ if } m \leq \mu_2}\vspace{0.2cm}\\
\displaystyle{\frac{1}{\hat\sigma_3}\exp\left[-\frac{(m-\mu_2)^2}{2\hat\sigma_3^2}\right] \mbox{ if } m > \mu_2}.
\end{array}\right.
\label{eq:asymGauss}
\]
\item{} The combinatorial background $\mathcal{F}_{c}$, arising from random combinations of real $J/\psi$ mesons with additional tracks, is modelled by a first-order polynomial
\[
\mathcal{F}_{c}(m | p_0, p_1) = p_0 + p_1 m .
\]
\end{itemize}
All the parameters in Eq.~(\ref{eq:pionBkg}) ($\mu_1,\mu_2,\hat\sigma_1,\hat\sigma_2,\hat\sigma_3,\gamma$) are obtained from fits to MC simulations, performed using \PYTHIA[8]~\textsc{A14-rb}, of $B^\pm \to J/\psi \pi^\pm$ in bins of the jet $\pt$ and independently of $z$ or $\ptrel$. The normalisation of this background, $\lambda_{B_\pi}$, is fixed relative to the signal as the ratio $\mathcal{R}(\pi,K)$ of the branching fractions of signal and background~\cite{pdg2018}, i.e.
\[
\frac{\lambda_{B_\pi}}{\lambda_s} = \mathcal{R}(\pi,K) = \frac{\mathcal{B}(B^\pm \rightarrow J/\psi \pi^\pm)}{\mathcal{B}(B^\pm \rightarrow J/\psi K^\pm)} = 3.84\%.
\]
The fits are performed individually for each $(\pt,z)$ and $(\pt,\ptrel)$ bin in the mass range $[5.0, 5.7]~\GeV$, and each of them has ten free parameters: four for the signal $(\mu, \sigma_1, \sigma_2, \beta)$, two for the misreconstructed background $(b,s)$, two for the combinatorial background $(p_0, p_1)$, and two for the free normalisations of each component, $\lambda_i$. Figure~\ref{fig:fits} shows the results of such fits in four representative bins of the longitudinal profile $z$ and transverse profile $\ptrel$ for the lowest and highest jet-$\pt$ selections.\\
\newline
The quality of the fits is quantified using $\chi^2$ functions, defined as $\chi^2 = \sum_i\left[(x_i-f_i)^2 / (\Delta x_i)^2\right]$ for each fit. Here, $x_i$ is the value of the data distribution in bin $i$; $\Delta x_i$ is its statistical uncertainty, and $f_i$ is the corresponding value of the fit model. The values of the $\chi^2$ divided by the number of degrees of freedom, $N_{\mathrm{dof}} =30$, range from 0.6 to 2.2. The fit results are studied as a function of the jet $\pt$ and of the fragmentation variables $z$ and $\ptrel$. The results show that the signal component dominates the fit models with an average, global purity close to 69\%, followed by the misreconstructed background, with a global value of about 15\% and the combinatorial background, amounting to 12\% of the sample. Finally, the $B^\pm \to J/\psi\pi^\pm$ background amounts to 3\% of the sample. As a general trend, the purity of the sample increases as a function of $z$, while it is relatively constant as a function of $\ptrel$, particularly at low $\pt$. Figure~\ref{fig:fitResults} shows the signal purity and the fractions of the different backgrounds resulting from the fit as a function of $z$ and $\ptrel$, for the lowest and highest jet-$\pt$ selections. The hatched bands include the sum in quadrature of the statistical and systematic uncertainties of the fit, as discussed in Section~\ref{sec:systematics}.
\clearpage
\begin{figure}[H]
\vspace{2.5cm}
\includegraphics[width=0.49\textwidth]{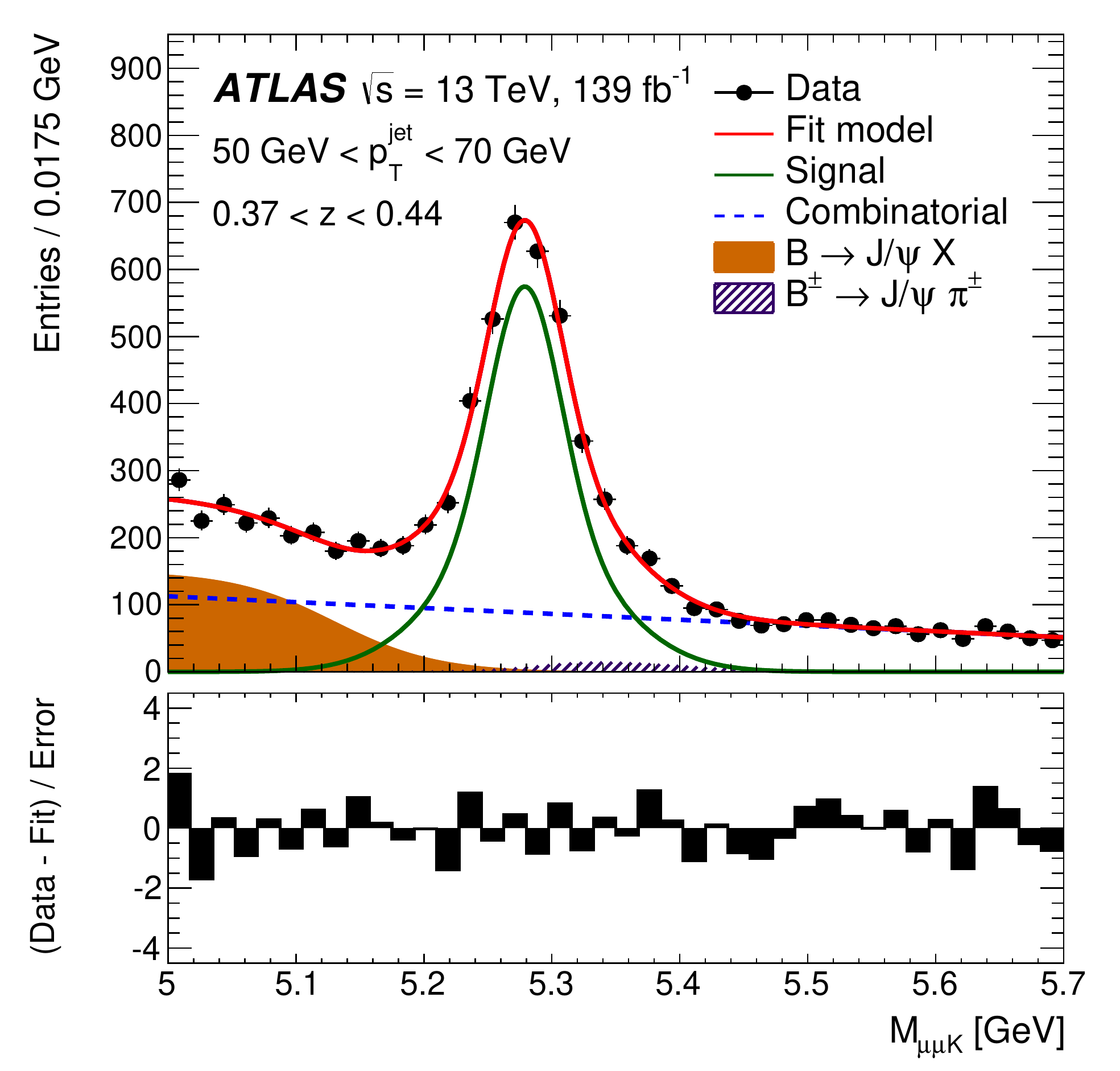}
\includegraphics[width=0.49\textwidth]{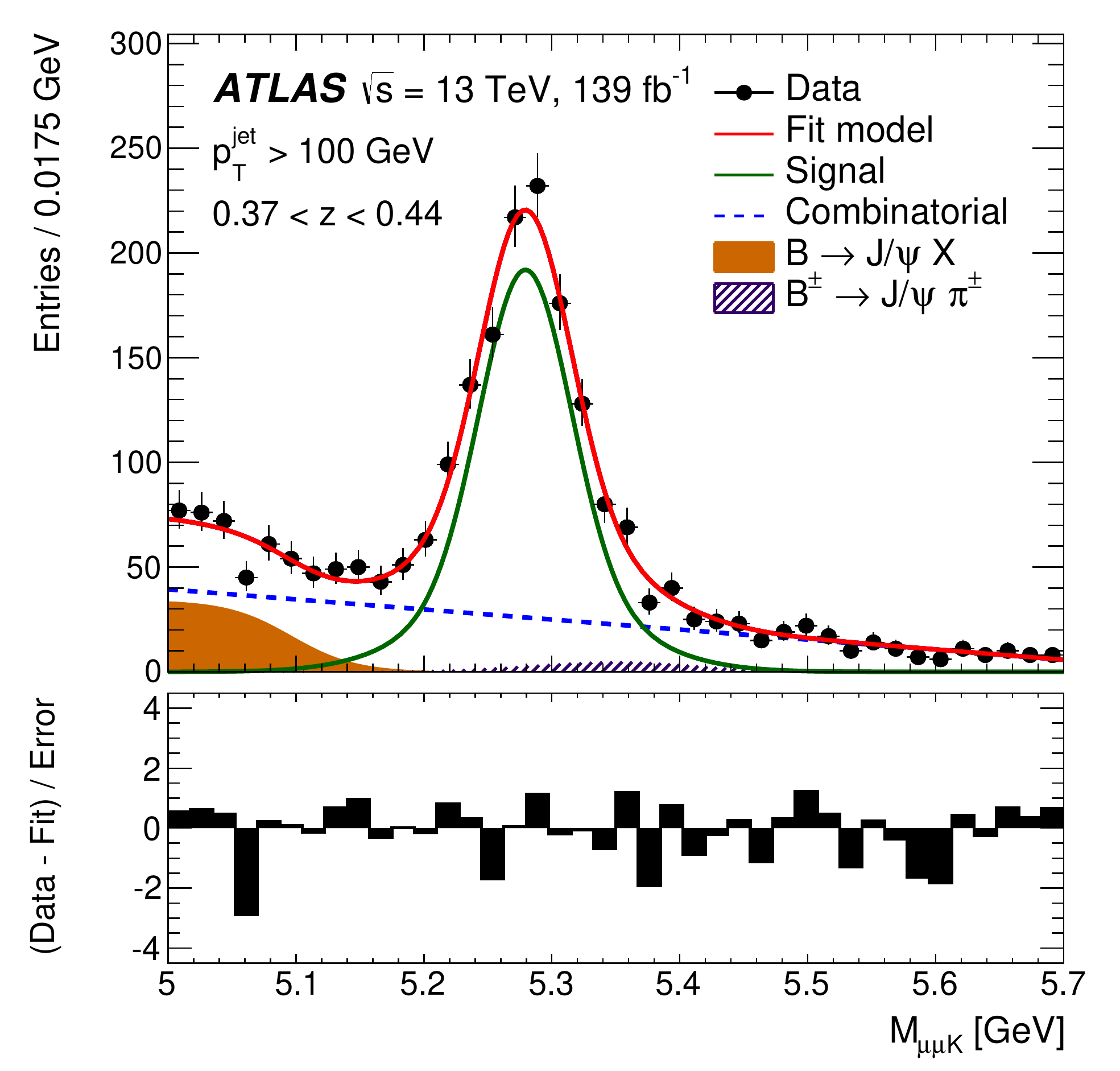}
 
\includegraphics[width=0.49\textwidth]{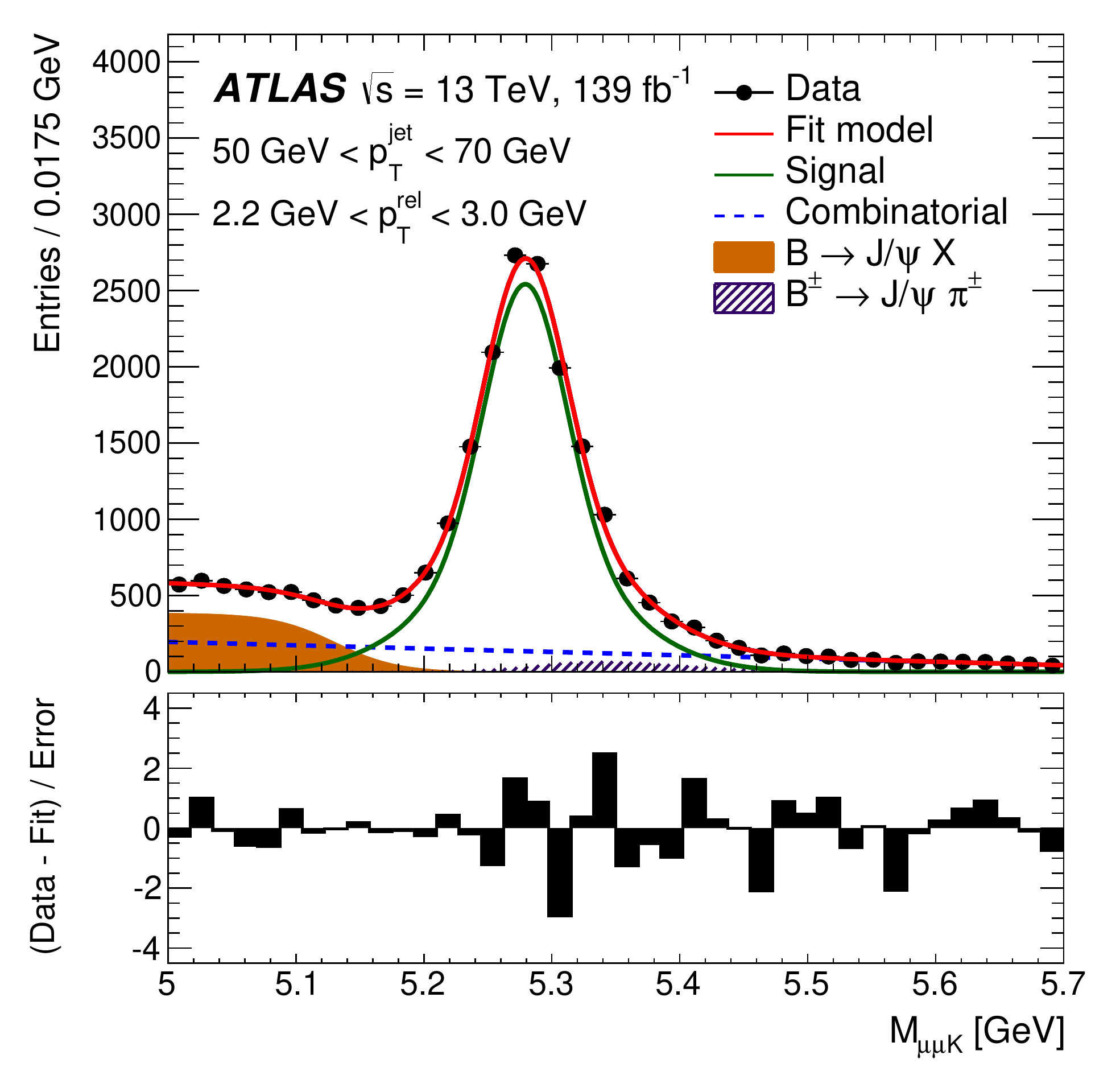}
\includegraphics[width=0.49\textwidth]{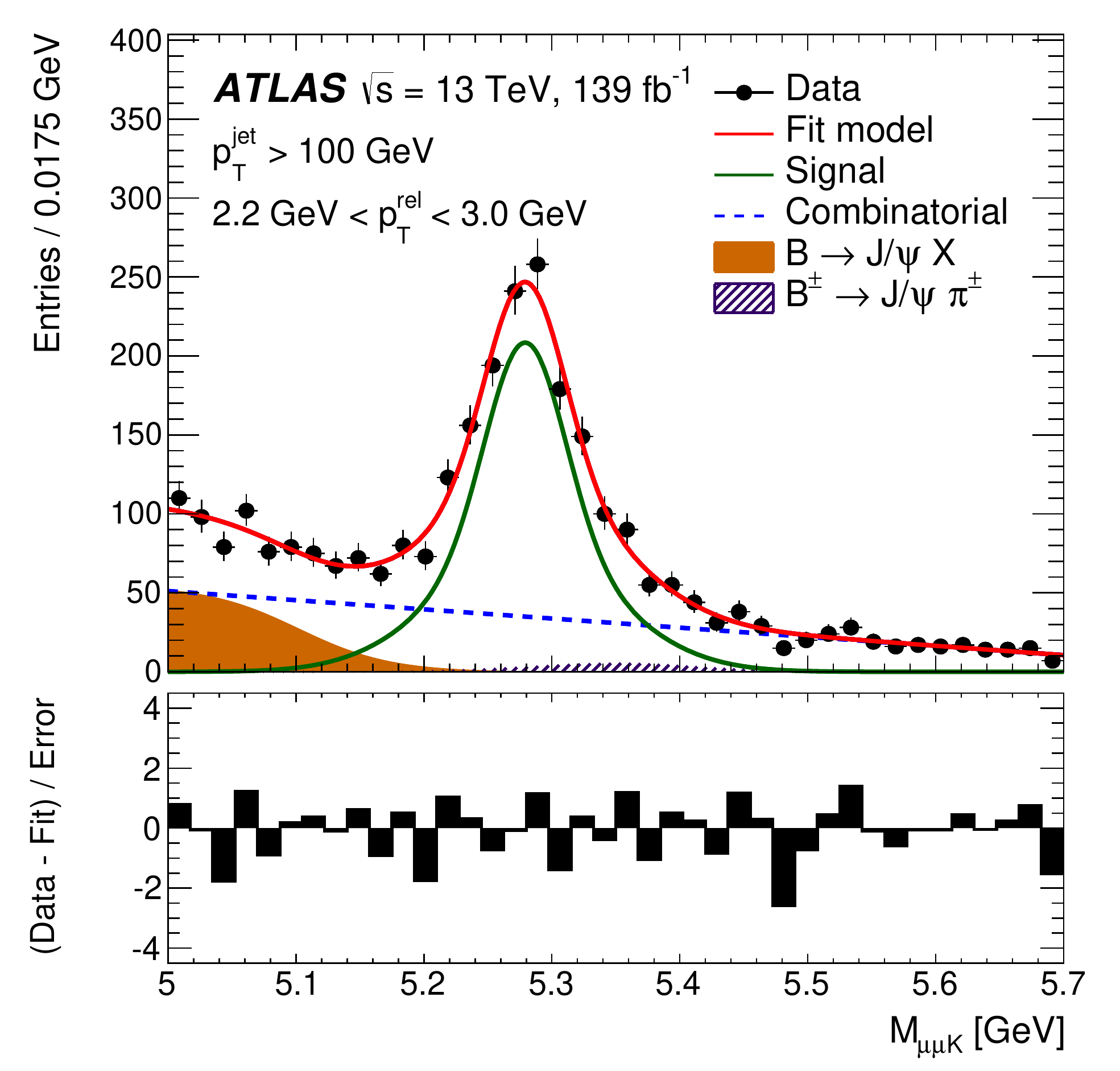}
\caption{Fits to the invariant mass distributions of $B^\pm$ candidates for $0.37 < z < 0.44$ in the lowest and highest jet-$\pt$ bins (top) and $2.2~\GeV < \ptrel < 3.0~\GeV$ in the lowest and highest jet-$\pt$ bins (bottom). The bottom panels show the difference between the data and the fit, divided by the statistical uncertainty of the data.}
\label{fig:fits}
\end{figure}
\clearpage
\begin{figure}[H]
\vspace{1.6cm}
\includegraphics[width=0.49\textwidth]{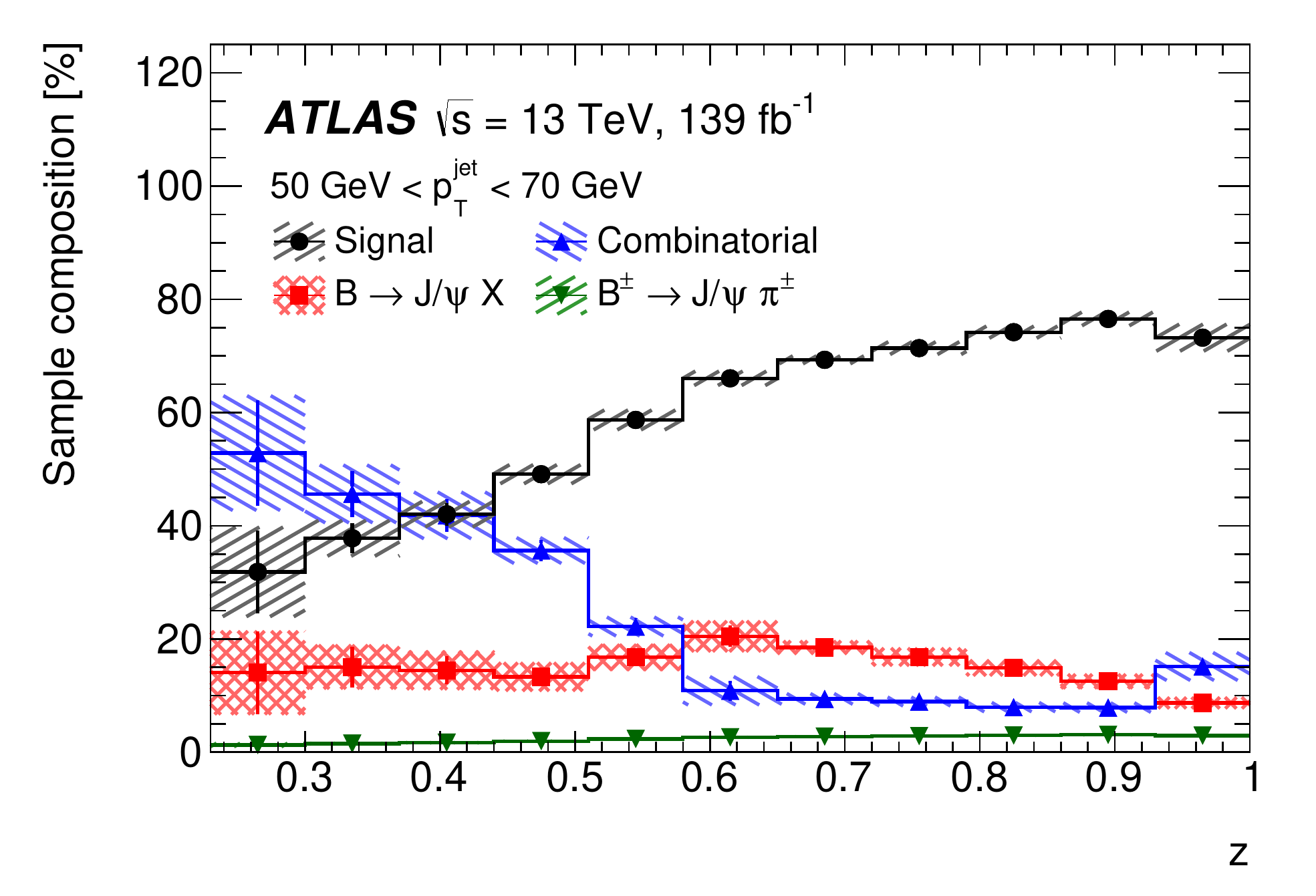}
\includegraphics[width=0.49\textwidth]{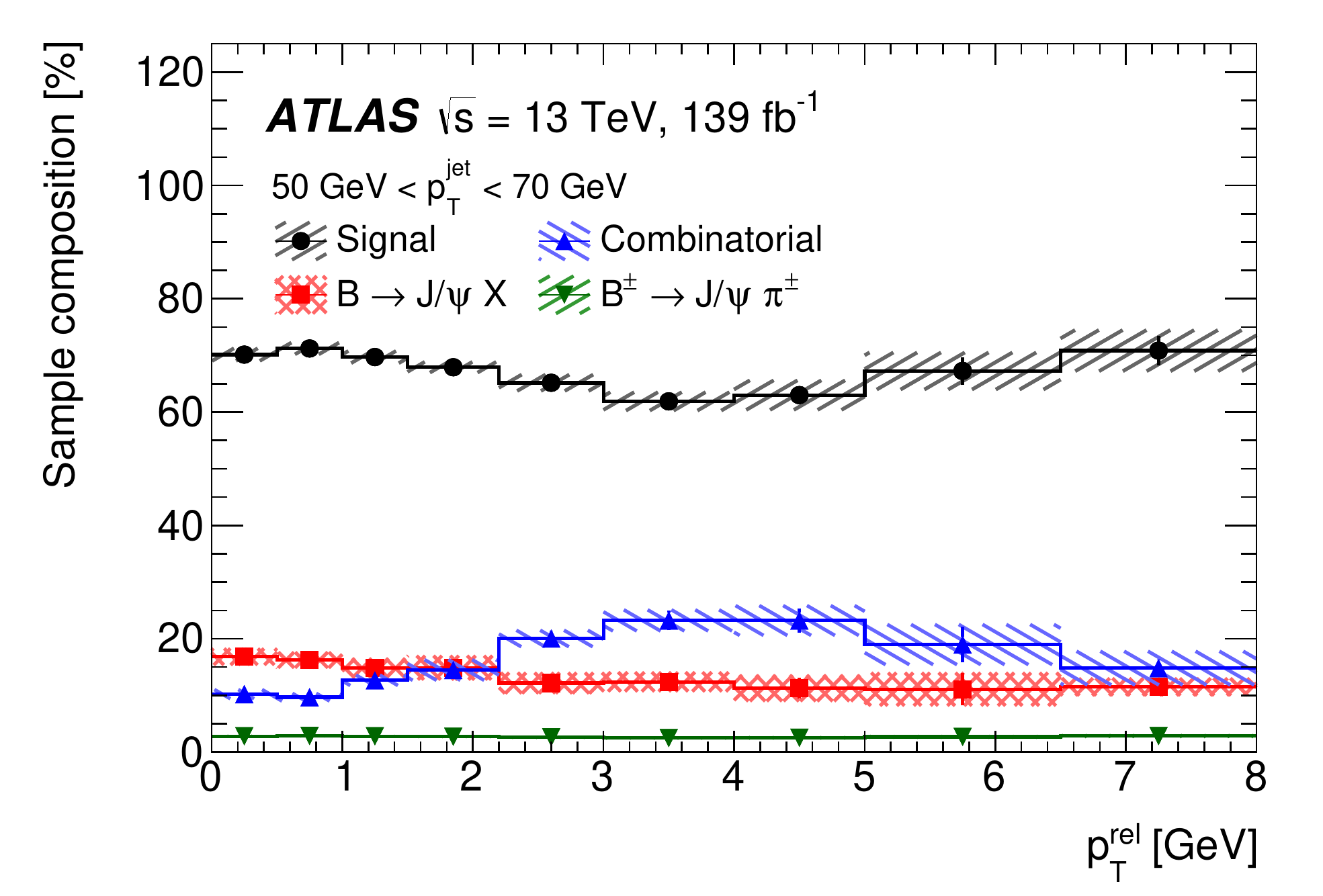}
 
\includegraphics[width=0.49\textwidth]{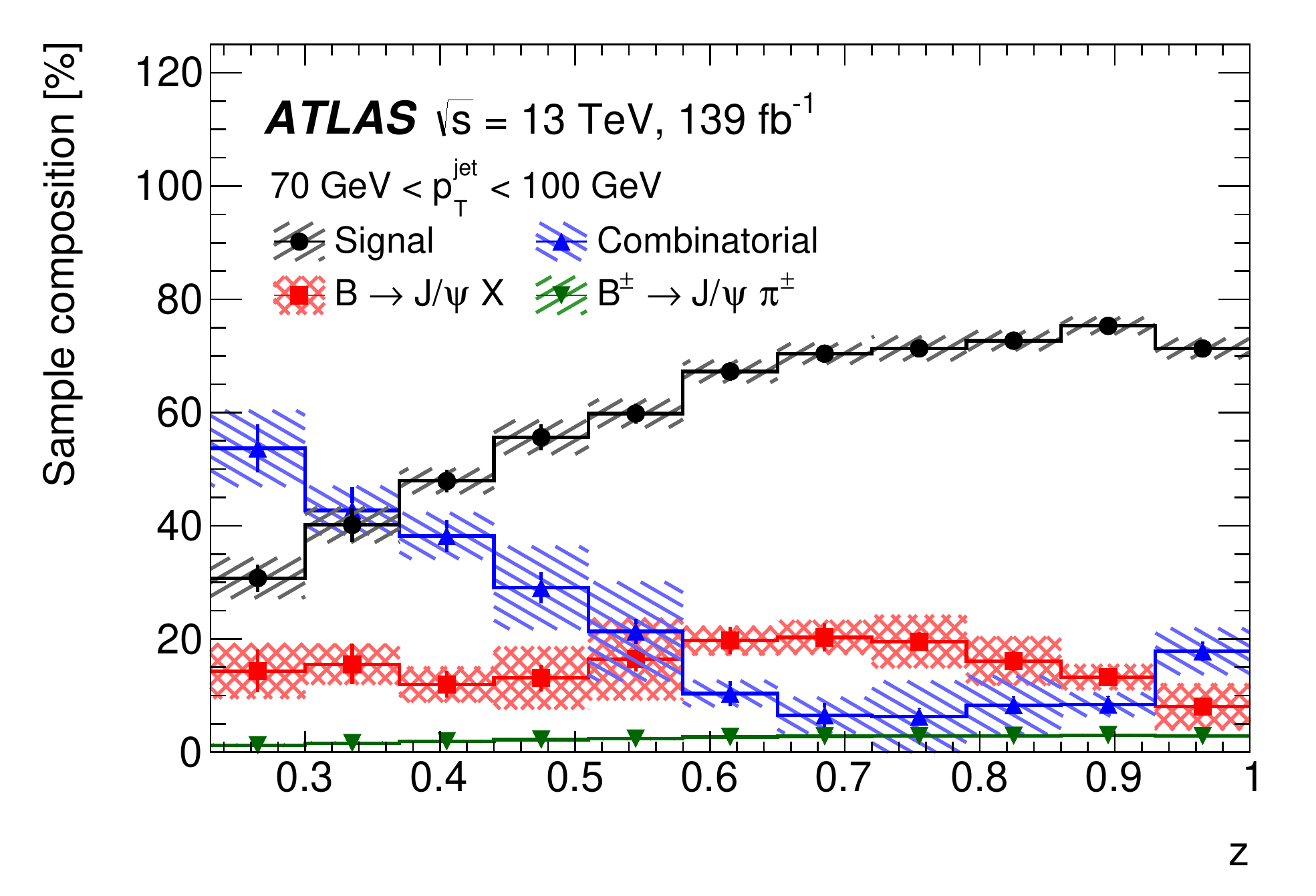}
\includegraphics[width=0.49\textwidth]{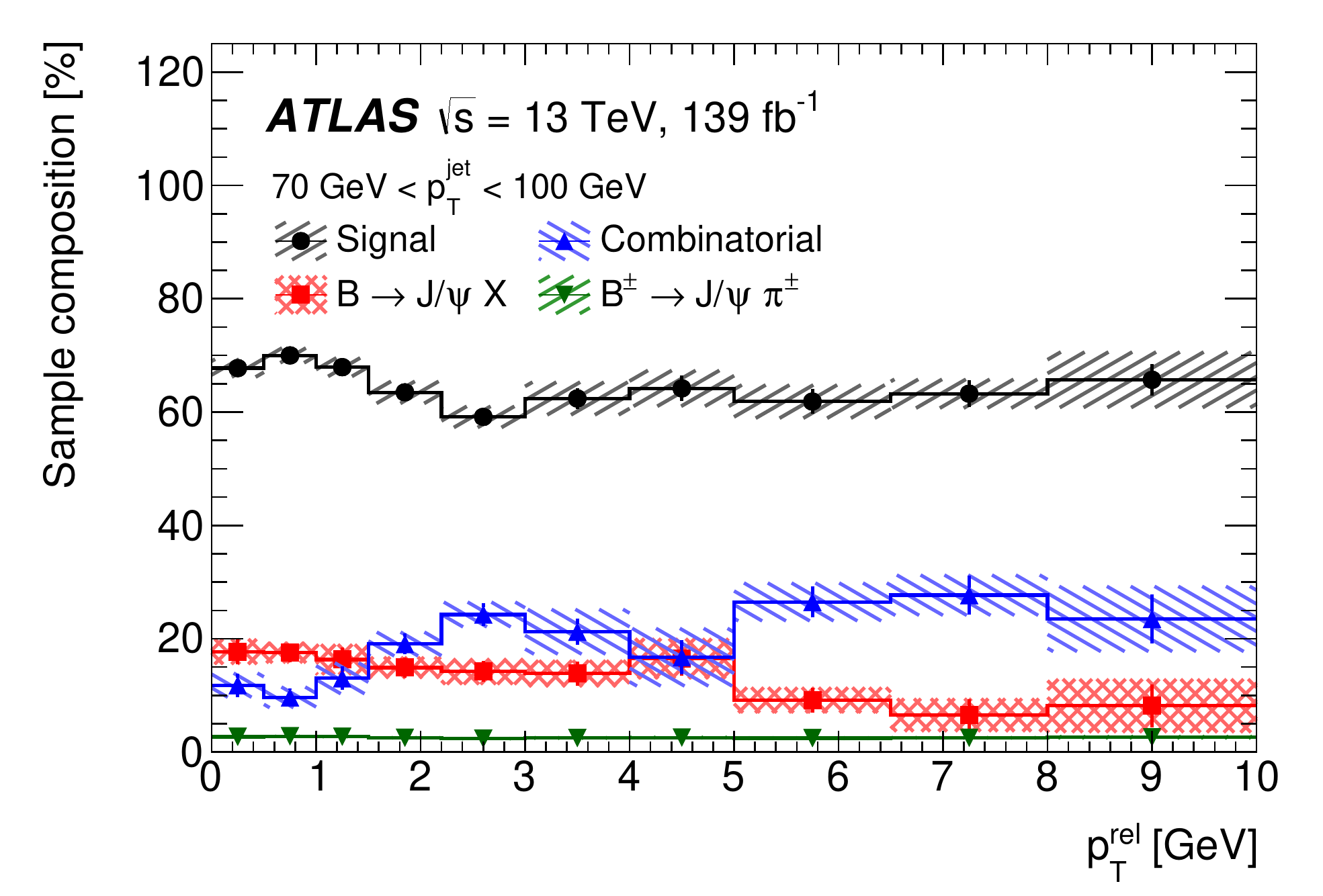}
 
\includegraphics[width=0.49\textwidth]{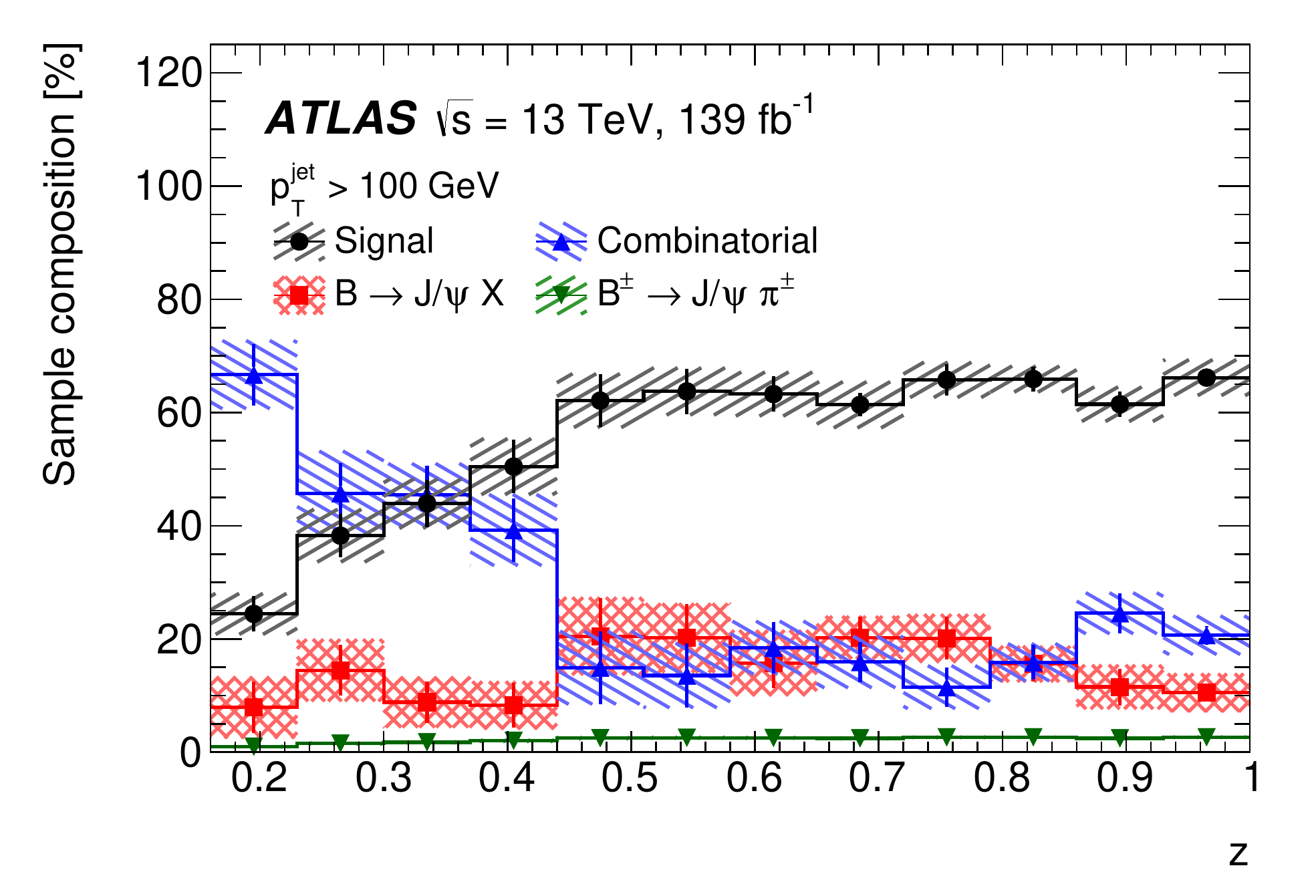}
\includegraphics[width=0.49\textwidth]{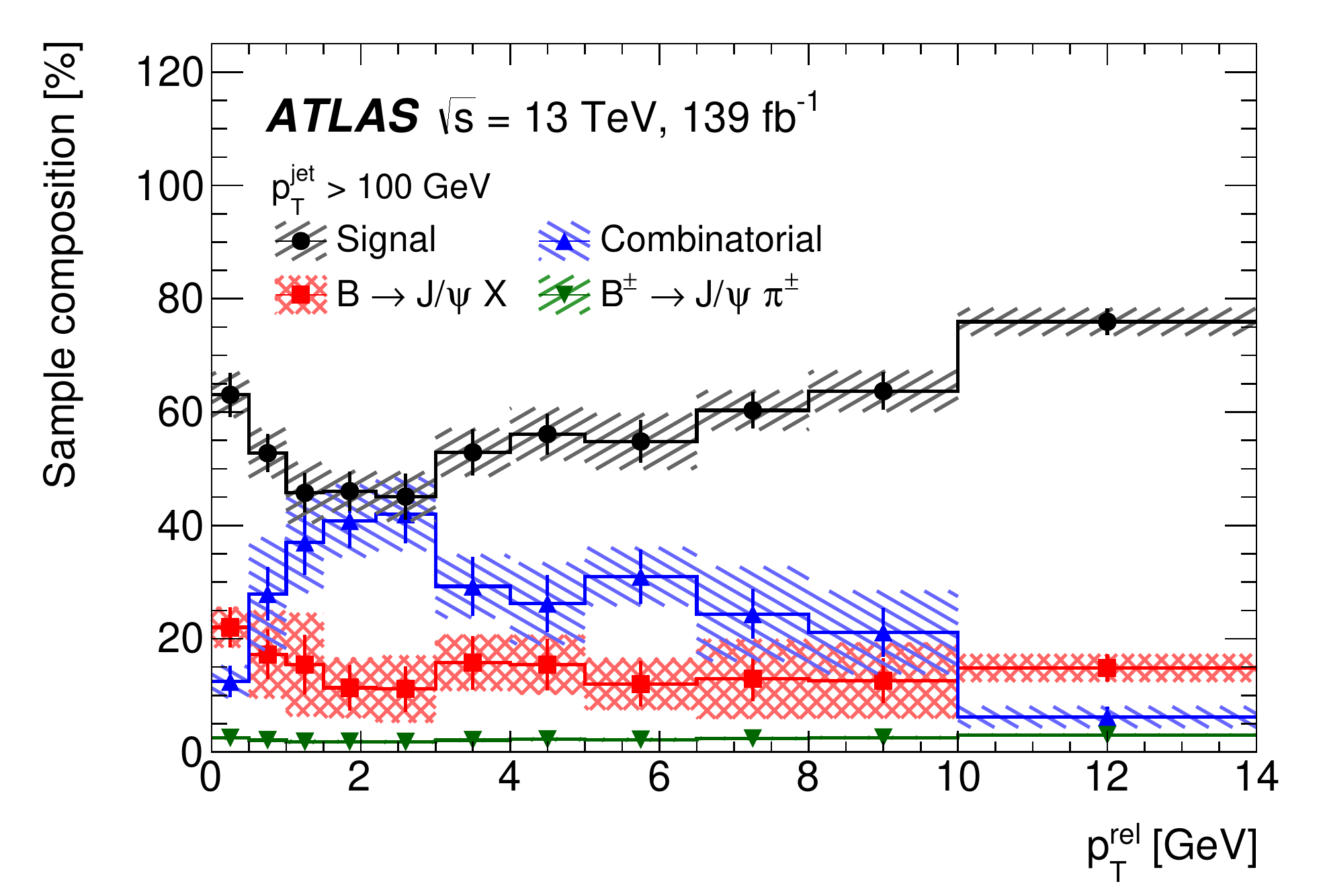}
\caption{Fit results as a function of $z$ for all jet-$\pt$ bins (left) and as a function of $\ptrel$ (right) for all jet-$\pt$ bins. The black circular markers represent the purity $p(\pt,x)$ for each bin. The error bars represent the statistical uncertainty of the fitted values, while the hatched bands represent the sum in quadrature of the statistical and systematic uncertainties of the fits.}
\label{fig:fitResults}
\end{figure}
\clearpage
% End of text imported from the .//purity.tex input file
\section{Unfolding to particle level}
\label{sec:unfolding}
% The next lines are included from the .//unfolding.tex input file
The detector-level distributions, from which the background processes have been subtracted following the methodology described in Section~\ref{sec:purity}, are corrected for detector effects such as inefficiencies and migrations caused by the finite resolution of the detector. The particle-level observables are defined using requirements equivalent to the detector-level selection described in Section~\ref{sec:data-selection}. Candidate $B^{\pm}\to J/\psi K^{\pm}$ decays are selected if the muons from the $J/\psi$ decay have $\pt > 6~\GeV$ and the kaon has $\pt > 4~\GeV$. The kinematics of the $B^\pm$ hadrons are reconstructed before QED radiation. Jets are reconstructed using the anti-$k_t$ algorithm with $R=0.4$, using all particles with average lifetime $\tau > 10$ ps, including muons and neutrinos. Jets with $\pt> 20~\GeV$ and $|\eta| < 2.1$ are selected, and those overlapping with another jet with $\pt > 20~\GeV$ within $\Delta R = 0.8$ are discarded. While the measurement is performed in the fiducial phase space with $\pt > 50~\GeV$, an underflow bin with $35~\GeV < \pt < 50~\GeV$ is used to take into account the migrations from phase space regions below the jet $\pt$ threshold.\footnote{Data falling in this bin are also corrected for background sources as described in Section~\ref{sec:purity}.} The $B^\pm$ mesons are matched to jets if the angular distance between them fulfils $\Delta R < 0.4$. Finally, the four-momenta of any decay products of the $B^\pm$ meson lying outside the jet ($\Delta R > 0.4$) are added to the jet four-momentum.\\
\newline
The unfolding procedure can be parameterised using a transfer matrix $A_{ij}$, which contains the information about the bin-by-bin migrations from the detector-level distribution to the particle-level distributions, and is written as
\[
R_i = \sum_{j=1}^N \frac{\mathcal{E}_j}{\mathcal{P}_i} A_{ij} T_j.
\]
The unfolding can thus be regarded as a system of linear equations, for which the solution is the particle-level distribution $T_j$. The transfer matrix $A_{ij}$, the efficiency $\mathcal{E}_j$ and the purity\footnote{This purity is unrelated to the purity described in Section~\ref{sec:purity} and corrected for before the unfolding.} $\mathcal{P}_i$ are determined using the \PYTHIA[8] \textsc{A14-rb} MC sample, and $R_i$ are the values of the detector-level distribution. An iterative matrix inversion method based on Bayes' theorem~\cite{bayes} is used, as implemented in the \textsc{RooUnfold} program~\cite{roounfold}. The number of iterations is chosen so that the sum in quadrature of the statistical uncertainty and the uncertainty associated with the mismodelling in the unfolding, described in Section~\ref{sec:systematics}, is minimised. This results in four iterations for the unfolding of both the longitudinal and transverse profiles. The unfolding is performed in such a way that the migrations in both dimensions, between the fragmentation variable ($z$ or $\ptrel$) and the jet $\pt$ are taken into account. The correlations in the statistical uncertainties, as well as those between the particle-level and detector-level MC distributions, are treated by using pseudo-experiments~\cite{bootstrap} in the calculation of each component of the unfolding, including matrices, efficiencies and purities as well as in the detector-level data.\\
\newline
The efficiency $\mathcal{E}$ is estimated as the fraction of particle-level jets, associated with a $B$ meson, which are matched within $\Delta R = 0.4$ to a detector-level jet that contains a reconstructed $B$ meson. Its value ranges from 10\% to 30\% depending on $z$ and $\ptrel$. The efficiency decreases as a function of $z$ for all $\pt$ bins, although this behaviour is more pronounced for the higher $\pt$ bin. This is due to the fact that, in this boosted regime, the dimuon trigger cannot efficiently separate the trajectories of the two muons produced in the $J/\psi$ decay, and the events containing $B^\pm$ mesons are therefore not recorded efficiently. The values of $\mathcal{E}$ as a function of $\ptrel$ are significantly flatter.\\
\newline
The purity $\mathcal{P}$ is estimated as the fraction of detector-level jets which are matched to a particle-level jet within $\Delta R = 0.4$. Its value ranges from 60\% to 100\%, systematically increasing as a function of both $z$ and the jet $\pt$. As a function of $\ptrel$, the purity exhibits a decrease in the low $\pt$ region. This is due to the fact that reconstructed mesons flying in the same direction as the jet, as well as hadrons with high $\pt$, are more likely to arise from a true $B$ meson. In the high $\pt$ region, $\mathcal{P}$ is constant as a function of $\ptrel$.\\
\newline
The values of the matrix elements $A_{ij}$ are estimated using particle-level $B$-meson jets, geometrically matched to the corresponding detector-level objects within $\Delta R = 0.4$. Figure~\ref{fig:matrices} shows the two-dimensional transfer matrices as a function of $(\pt, z)$ and $(\pt,\ptrel)$.
\begin{figure}[H]
\centering
\includegraphics[width=0.70\textwidth]{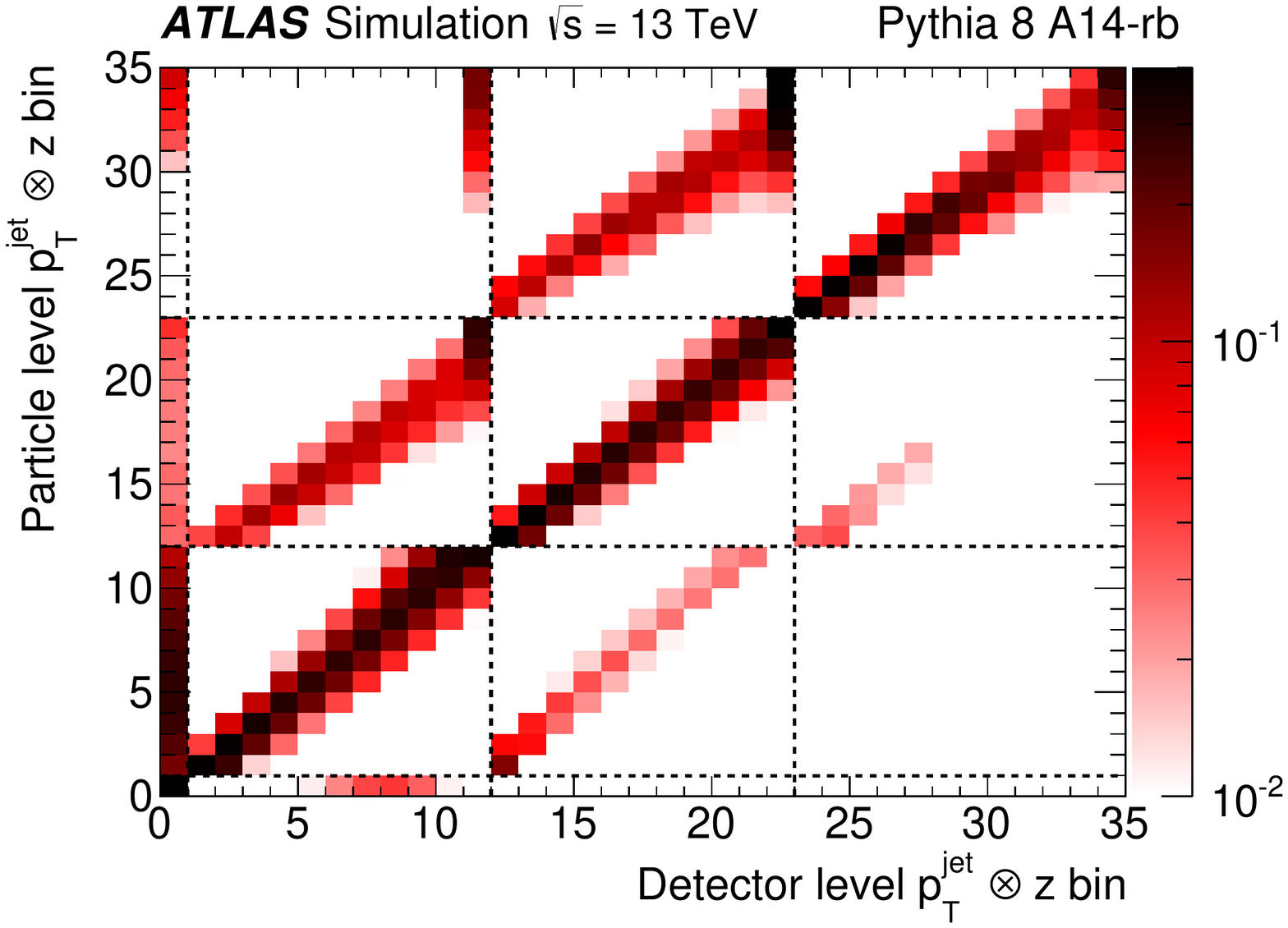}\\
\includegraphics[width=0.70\textwidth]{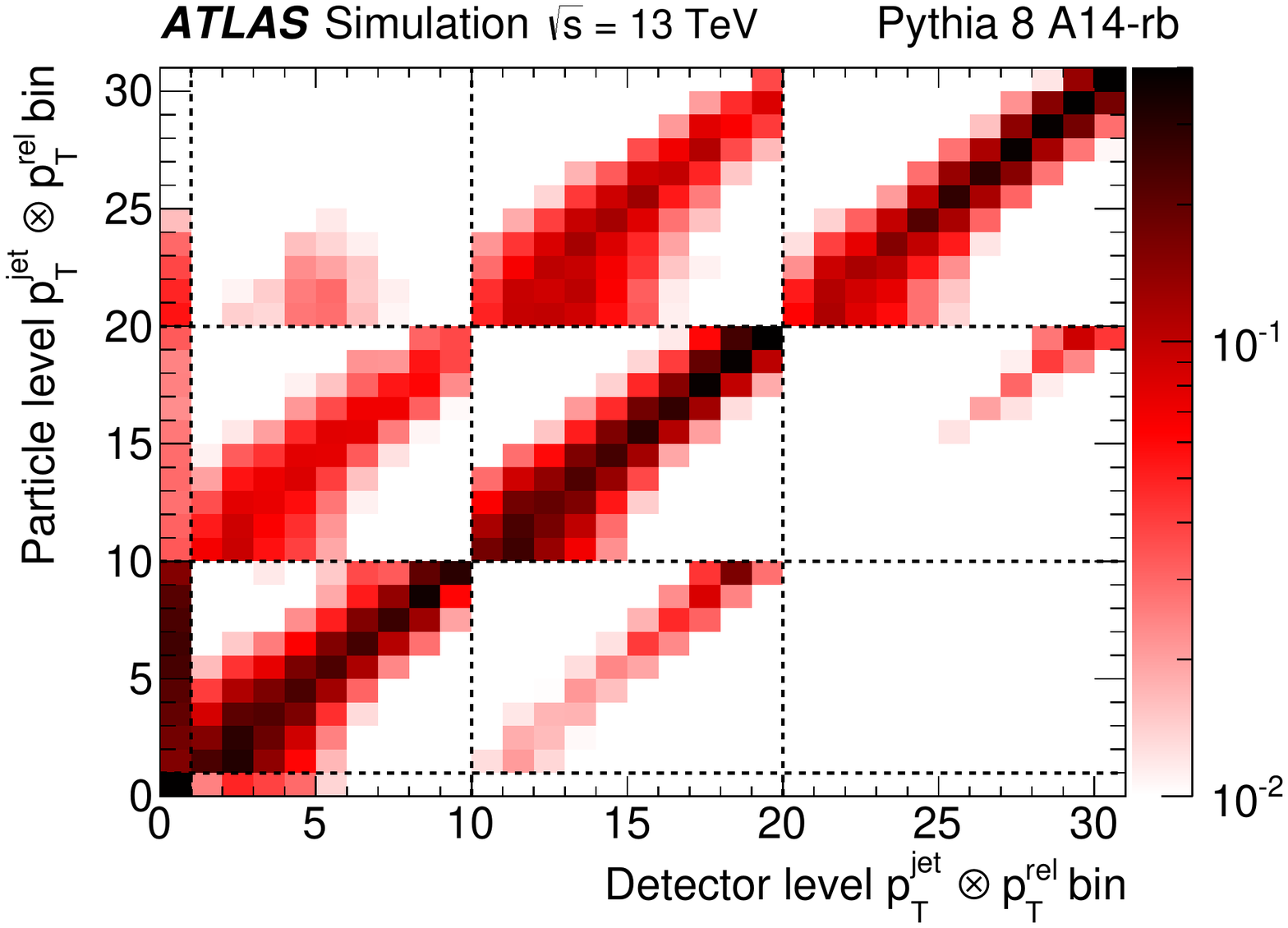}
\caption{Transfer matrices, obtained using the \PYTHIA[8] \textsc{A14-rb} MC sample, as a function of the jet $\pt$ and the longitudinal profile $z$ (top) and the transverse profile $\ptrel$ (bottom). Each small square represents a bin in $(\pt, z)$ or $(\pt, \ptrel)$, while each large box represents one of the three jet-$\pt$ intervals used in the analysis. The first bin, containing jets with $\pt < 50 ~\GeV$, accounts for the underflow.}
\label{fig:matrices}
\end{figure}
 
% End of text imported from the .//unfolding.tex input file
\clearpage
\section{Systematic uncertainties}
\label{sec:systematics}
% The next lines are included from the .//systematics.tex input file
The systematic uncertainties in the measurement are classified into four main categories. The first of them concerns the identification of $B^{\pm}$ mesons, and includes the systematic uncertainties in muon reconstruction as well as the uncertainties in the fit procedure for the determination of the signal purity described in Section~\ref{sec:purity}; the second category concerns jet reconstruction and identification, including the jet energy scale and resolution, the jet angular resolution and the efficiencies of the JVT algorithm. The third category is related to the unfolding procedure, including both the uncertainty in the MC model used for the corrections and the uncertainty due to mismodelling of the data by the nominal MC simulation. The fourth category takes into account systematic effects in the MC description of the pile-up dependence of the measurement.\\
\newline
The systematic uncertainties from all experimental sources are estimated by using the detector-level MC events. The resulting shifted distributions are then corrected for detector effects following the unfolding procedure described in Section~\ref{sec:unfolding}, normalised to unit area, and compared with the nominal MC distributions, also normalised to unit area. The results of this comparison constitute the systematic uncertainties of the unfolded results for the normalised differential cross sections, and exploit the correlations between the numerator and the denominator when applying the normalisation. The full set of systematic uncertainties are described below, together with a detailed description of how they are estimated.
 
\paragraph{$B$ meson reconstruction uncertainties}
\begin{itemize}
\item{} The uncertainties in the purity corrections for the $B^\pm$ candidates are evaluated by considering other fit models as alternatives to those described in Section~\ref{sec:purity}. The probability density function for each component of the fit is substituted by an alternative function. For the signal model, one of the Gaussian functions in the double-Gaussian pdf is replaced by an asymmetric Gaussian function. The hyperbolic tangent shape in the misreconstructed-background pdf, $1-\tanh(x)$, is replaced by $1-\mathrm{erf}(x) = (2/\sqrt{\pi})\int_x^\infty \mathrm{e}^{-t^2}\mathrm{d}t$. For the $B^\pm \to J/\psi \pi^\pm$ background, one of the Gaussian functions in the double-Gaussian shape is replaced by a Crystal Ball function. Finally, the first-order polynomial describing the shape of the combinatorial background is replaced by an exponential function. The total uncertainty in the purity corrections is defined as the sum in quadrature of the deviations of the varied fit results, evaluated separately, from the nominal fit model described in Section~\ref{sec:purity}. The impact of this uncertainty ranges from 1\% to a maximum of 17\% for the measurement of $z$, and from 1\% to 8\% for the measurement of $\ptrel$, systematically increasing with the jet $\pt$.
 
\item{} Muon momentum scale and resolution uncertainties~\cite{muons} include variations in the smearing of the inner detector and muon spectrometer tracks, variations of the muon momentum scale, and additional charge-dependent variations related to the correction for the sagitta bias. These variations are applied to the muon selection and both when adding the muon four-momenta to the jets using the ghost-association method described in Section~\ref{sec:data-selection} and when reconstructing the mass and four-momenta of the $B^{\pm}$ mesons. These are applied in a fully correlated way. The impact of this uncertainty is below 2\% in all regions of the phase space, for both the $z$ and $\ptrel$ measurements.
 
\item{} Muon identification uncertainties~\cite{MUON-2018-03} cover variations of the muon efficiency corrections described in Section~\ref{sec:data-selection}. The variations include $1\sigma$ shifts of the statistical and systematic uncertainties in the efficiency corrections. As in the muon reconstruction case, the variations are applied, in a fully correlated way, both when adding muons to the jets and when reconstructing the $B$ hadron momenta. The impact of this uncertainty is small, with values below 3\% in all regions of the phase space.
 
\item{} The uncertainty in the dimuon trigger efficiency~\cite{TRIG-2018-01} is evaluated by using correction factors to take into account the uncertainty in the mismodelling of the trigger efficiency. These correction factors are applied as per-jet weights depending on the $\pt$ and rapidity of the muons, as well as the angular distance $\Delta R$ between the two muons in the $B^\pm$ decay. The difference between the weighted distribution and the default one defines the systematic uncertainty. The values of this uncertainty range from a few per mille to a maximum of 4\% in the lower tails of the $z$ distribution, and from a few per mille to a maximum of 2\% as a function of $\ptrel$.
 
\item{} The uncertainty in the kaon reconstruction efficiency is estimated by randomly rejecting 2\% of the kaon tracks from the reconstruction, following a uniform random distribution~\cite{PERF-2015-08}. The impact on the fragmentation properties is at the per mille level for both the $z$ and $\ptrel$ measurements.
\end{itemize}
 
\paragraph{Jet-related uncertainties}
\begin{itemize}
\item{} The jet energy scale (JES) and jet energy resolution (JER) uncertainties are estimated as described in Ref.~\cite{JETM-2018-05}. The JES is calibrated on the basis of the simulation and in situ corrections obtained from data. The JES uncertainties are estimated using a correlation scheme comprising a set of 29 independent components, which depend on the jet $\pt$ and $\eta$. The total JES uncertainty in the $\pt$ of individual jets ranges from 2\% at $\pt = 50~\GeV$ to approximately 1\% at $\pt=100~\GeV$, with a mild dependence on $\eta$. The JER uncertainty is estimated using a correlation scheme involving 13 independent variations. Each of these variations ranges from about 0.8\% at $\pt=50~\GeV$ to about 0.5\% at $\pt = 100~\GeV$. In this measurement, the JES and JER uncertainties are propagated by varying the energy and $\pt$ of each jet by one standard deviation of each of the independent components. The impact of the JES uncertainty varies from a few per mille for low $\ptrel$ values at high jet-$\pt$ to 10\% for medium values of $z$ at low jet-$\pt$, and typically decreases with increasing jet $\pt$. The JER has a similar behaviour, varying from approximately 1\% for low $\ptrel$ values at high jet $\pt$ to 30\% at medium values of $z$ at low jet-$\pt$.
 
\item{} The impact of the jet angular resolution (JAR) uncertainty is estimated by smearing the angular coordinates $(\eta,\phi)$ of the jets by 10\% of the angular resolution of jets reconstructed from topological clusters~\cite{PERF-2014-07}, estimated in MC simulation. The $\eta$ and $\phi$ smearing is applied with the $\pt$ component of the jets held constant. The JAR uncertainty is negligible for the measurement of $z$. On the other hand, due to a linear dependence of $\ptrel$ on the sine of the angle between the $B$ meson and the jet, this uncertainty can have an impact of up to 3\% for the measurement of $\ptrel$.
 
\item{} The uncertainty from the efficiency corrections for the JVT algorithm described in Section~\ref{sec:data-selection}, used to mitigate the impact of pile-up jets, is evaluated by using alternative correction factors, shifted by the corresponding uncertainties. This variation has two components, each of which shifts the correction factors in opposite directions. The resulting uncertainty is very small, with values below 0.1\%.
\end{itemize}
 
\paragraph{Unfolding-related uncertainties}
\begin{itemize}
\item{} The effects of the mismodelling of the data by the MC simulation on the unfolding is accounted for as an additional source of uncertainty. This is assessed by reweighting the particle-level distributions so that the detector-level fragmentation variables predicted by the MC samples match those in the data. The modified detector-level distributions are then unfolded using the method described in Section~\ref{sec:unfolding}. The difference between the modified unfolded distribution and the reweighted particle-level distribution is taken as the uncertainty. The resulting uncertainty ranges from 0.2\% to 10\%, and typically increases with increasing values of $z$ and decreasing values of $\ptrel$.
\item{} The uncertainty due to using a particular MC model in the unfolding is estimated from the impact of two independent variations. The first accounts for the different amounts of gluon splitting in samples generated with different tunes. It is evaluated by comparing the results of the unfolding using the \textsc{A14} samples with those using the different gluon splitting fractions in the \Monash samples. These two particular tunes, which differ in the value of \alphas for final-state radiation,\footnote{The value of the strong coupling constant for time-like parton showers in the \textsc{A14} sample is $\alphas = 0.1260$, while the \Monash tune uses $\alphas = 0.1365$.} are chosen because fits to the gluon splitting fraction in data are compatible with the uncertainty band spanned by them. The second is related to the description of the measured observables made by different MC models, after disentangling the effect of different gluon splitting fractions. It is evaluated by repeating the unfolding using two alternative samples. While the nominal result is obtained using the \PYTHIA[8] \textsc{A14-rb} sample, the alternative samples are simulated using \PYTHIA[8] \textsc{A14} and \SHERPA, in which the fragmentation is simulated by the string model. To avoid double counting of the differences arising from the description of the gluon splitting, the \SHERPA and \PYTHIA[8] \textsc{A14} samples are reweighted so that the fraction of the selected jets arising from gluon splittings $g\to b\bar{b}$ is the same as in the \PYTHIA[8] \textsc{A14-rb} sample. The envelope of the differences between the results using \PYTHIA[8] \textsc{A14-rb} and both alternative samples covers the second source of systematic uncertainty. The total modelling uncertainty is defined by the sum in quadrature of these two variations. This uncertainty ranges from 4\% to 30\% and it is, together with the JES uncertainty, dominant for these measurements.
\end{itemize}
 
\paragraph{Pile-up-related uncertainties}
\begin{itemize}
\item{} The uncertainty due to pile-up is derived from the ratios of the unfolded results for low ($\mu < 32$) and high ($\mu \geq 32$) pile-up subsamples to the results for the nominal measurement. The uncertainty is defined as the envelope of the two ratios, and is below 10\% for both $z$ and $\ptrel$.
\end{itemize}
The total uncertainty is then determined as the sum in quadrature of the all the systematic uncertainties described above. The total uncertainty is larger for low values of the jet $\pt$, where it can reach values up to 30\% in some bins. For the higher $\pt$ bins, the uncertainty is systematically smaller, mainly because of the smaller jet energy scale components. Figure~\ref{fig:systematics} shows the values of the systematic uncertainties as a function of $z$ and $\ptrel$, for the lowest and highest bins of the jet $\pt$.\\
\newline
The dependence of the particle-level results on the choice of PDF was tested by using two alternative PDF sets for the \SHERPA predictions with the string fragmentation model. Predictions using \NNPDF[3.0] and \MMHT were compared with the nominal prediction, which uses \CT[14] (see Table~\ref{tab:mcSamples}). A maximum deviation of 2\% from the nominal results is observed. Moreover, the \CT[14] uncertainty, estimated using the full set of eigenvectors, produces a total variation of the order of a few per mille. The effect of varying the PDF is thus much smaller than the experimental uncertainties, and is therefore neglected.
\clearpage
\begin{figure}[H]
\vspace{2.cm}
\includegraphics[width=0.49\textwidth]{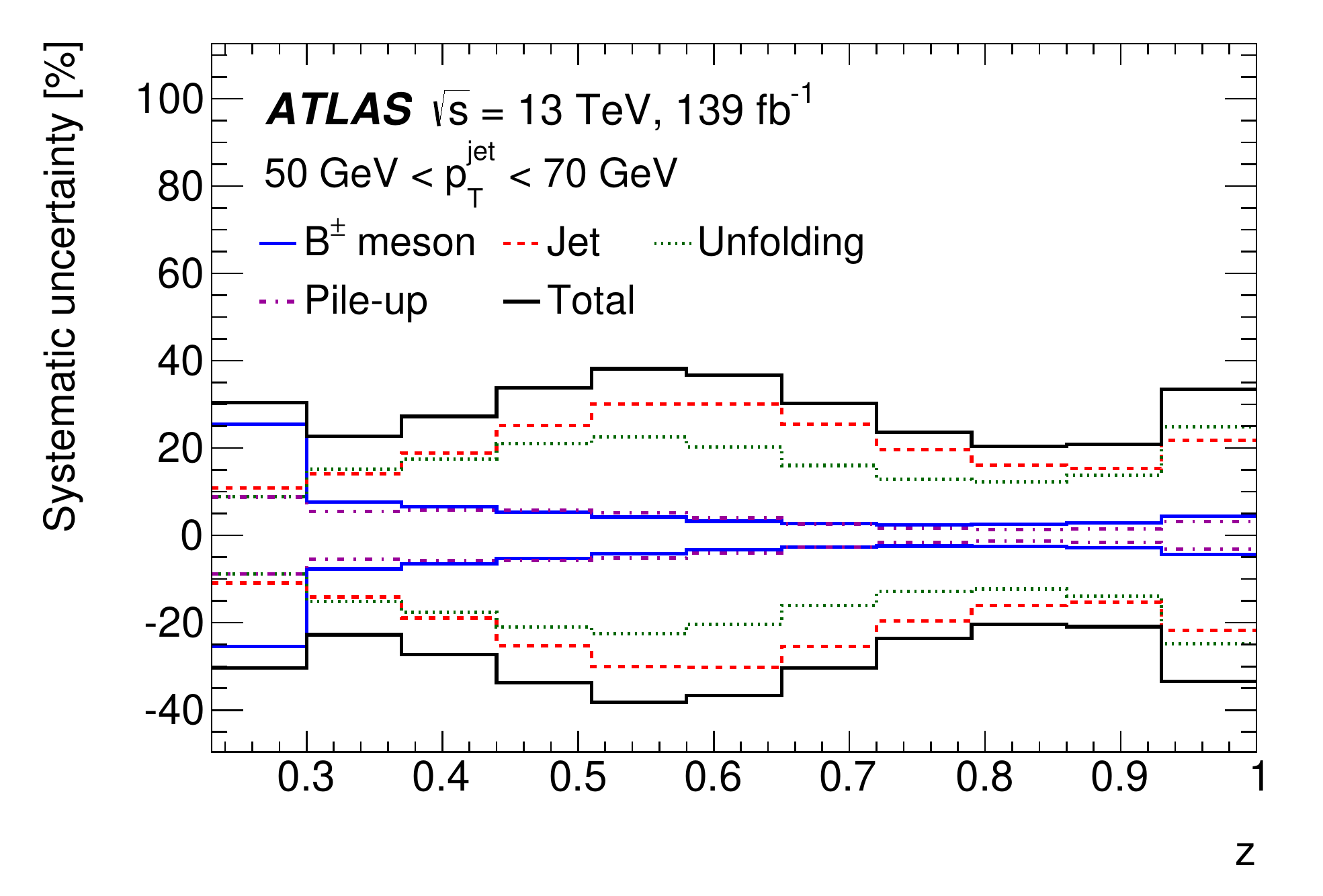}
\includegraphics[width=0.49\textwidth]{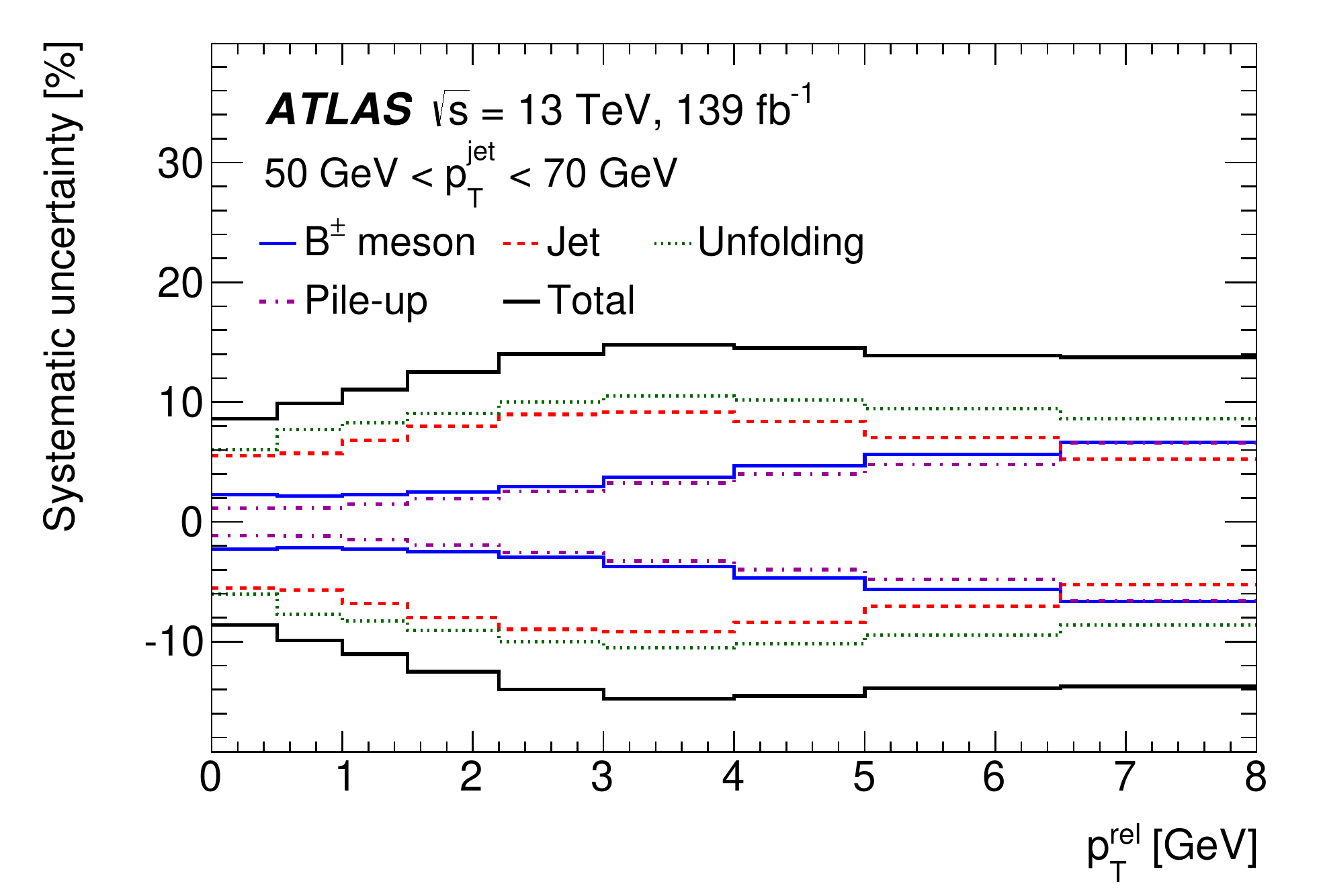}
 
\includegraphics[width=0.49\textwidth]{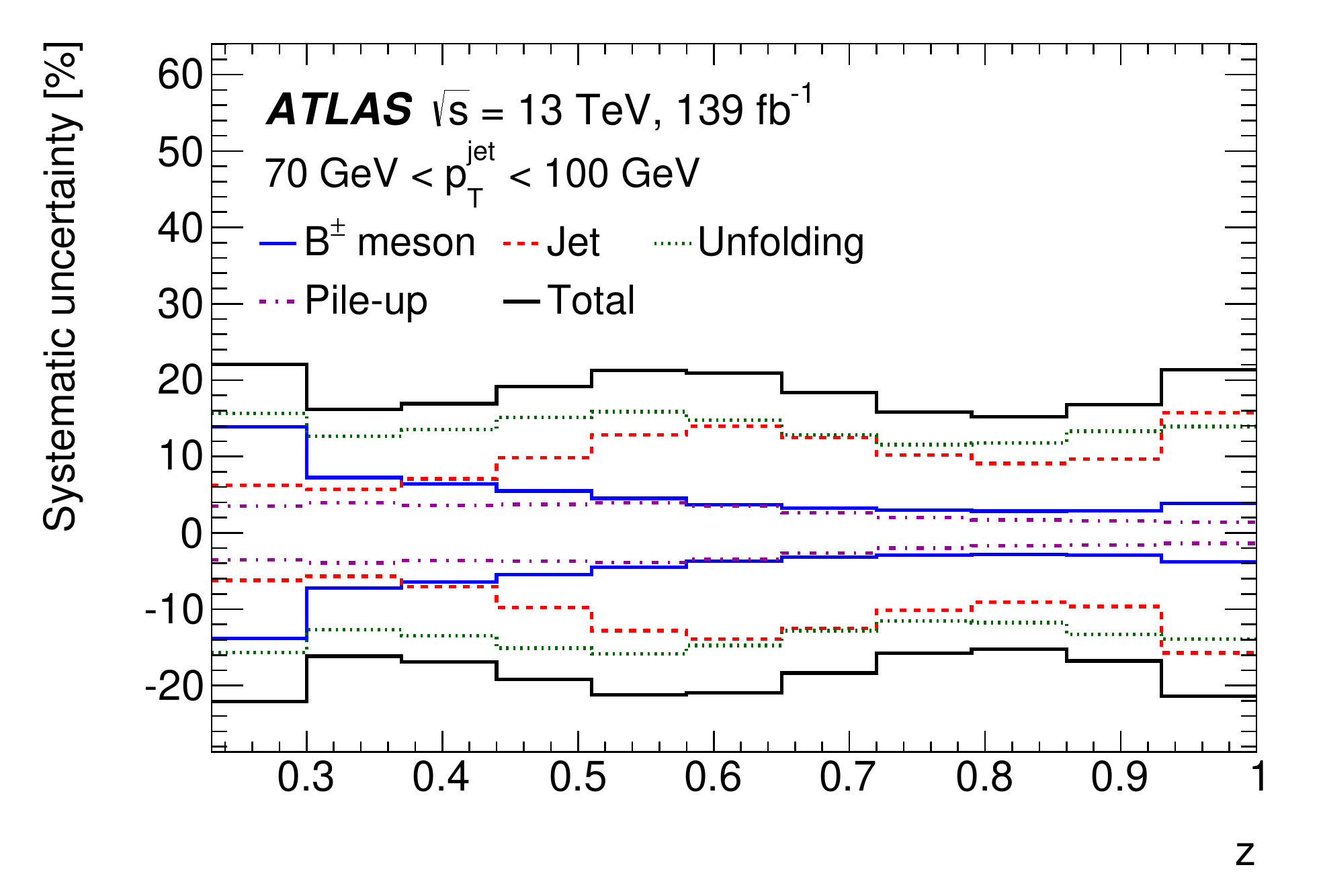}
\includegraphics[width=0.49\textwidth]{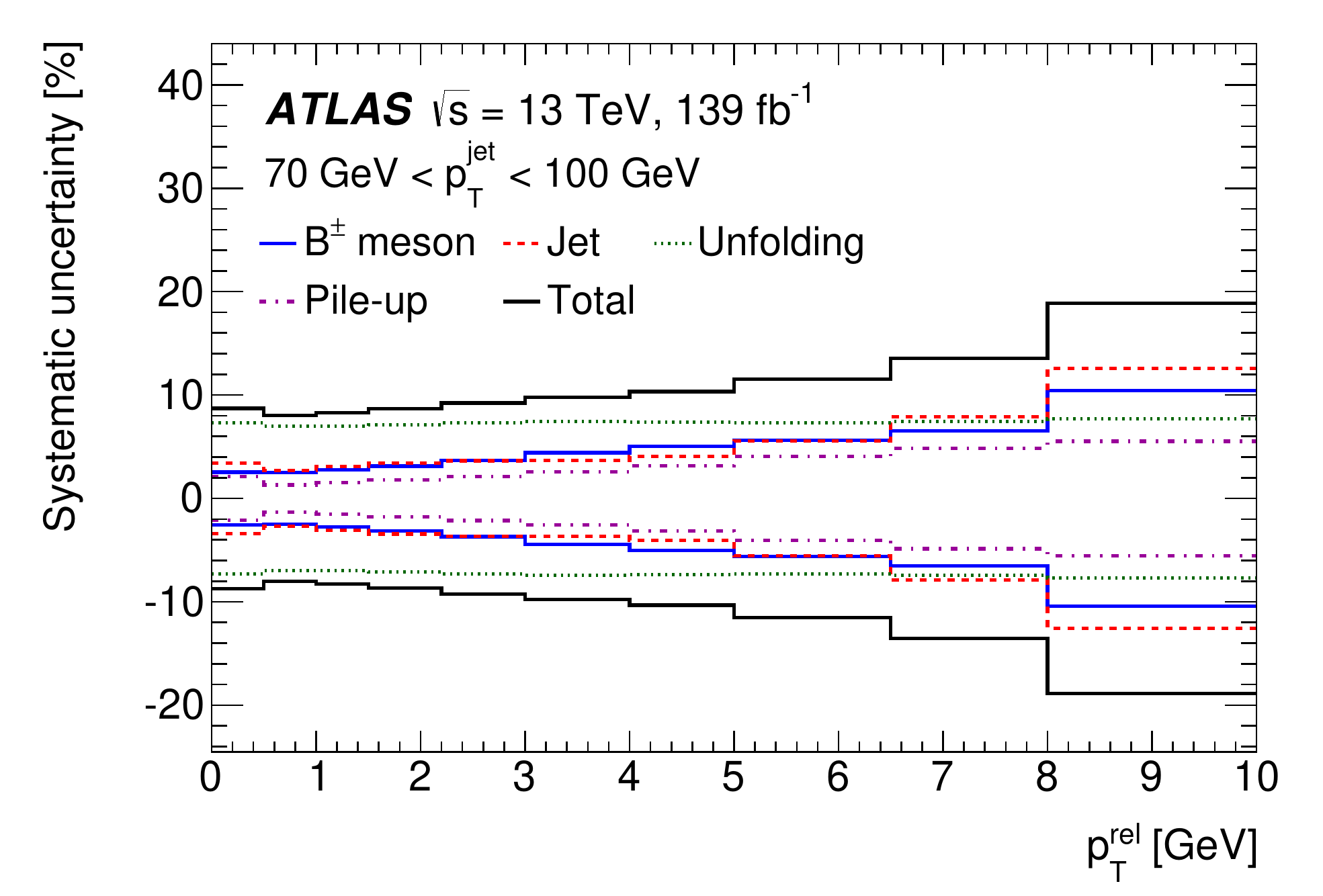}
 
\includegraphics[width=0.49\textwidth]{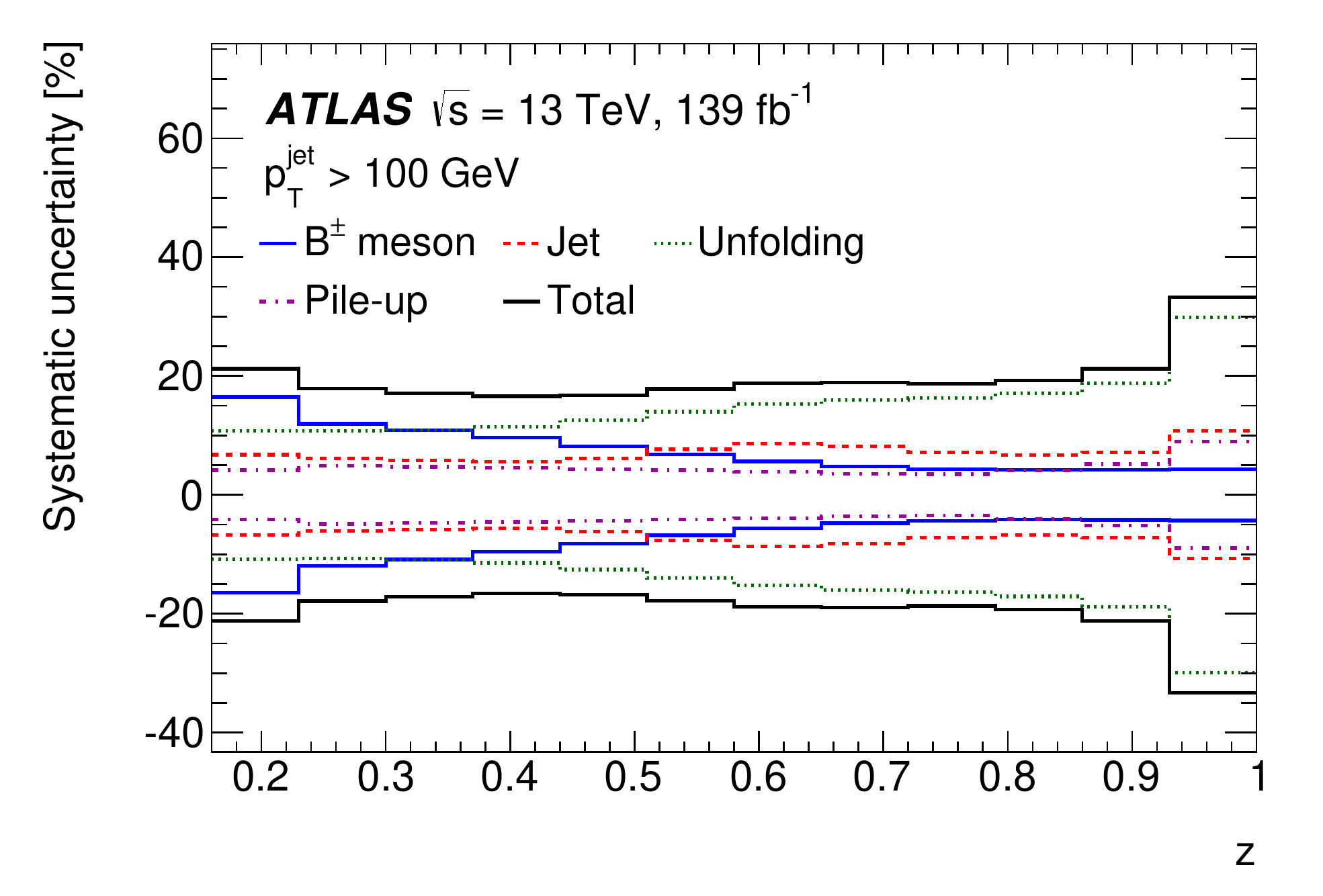}
\includegraphics[width=0.49\textwidth]{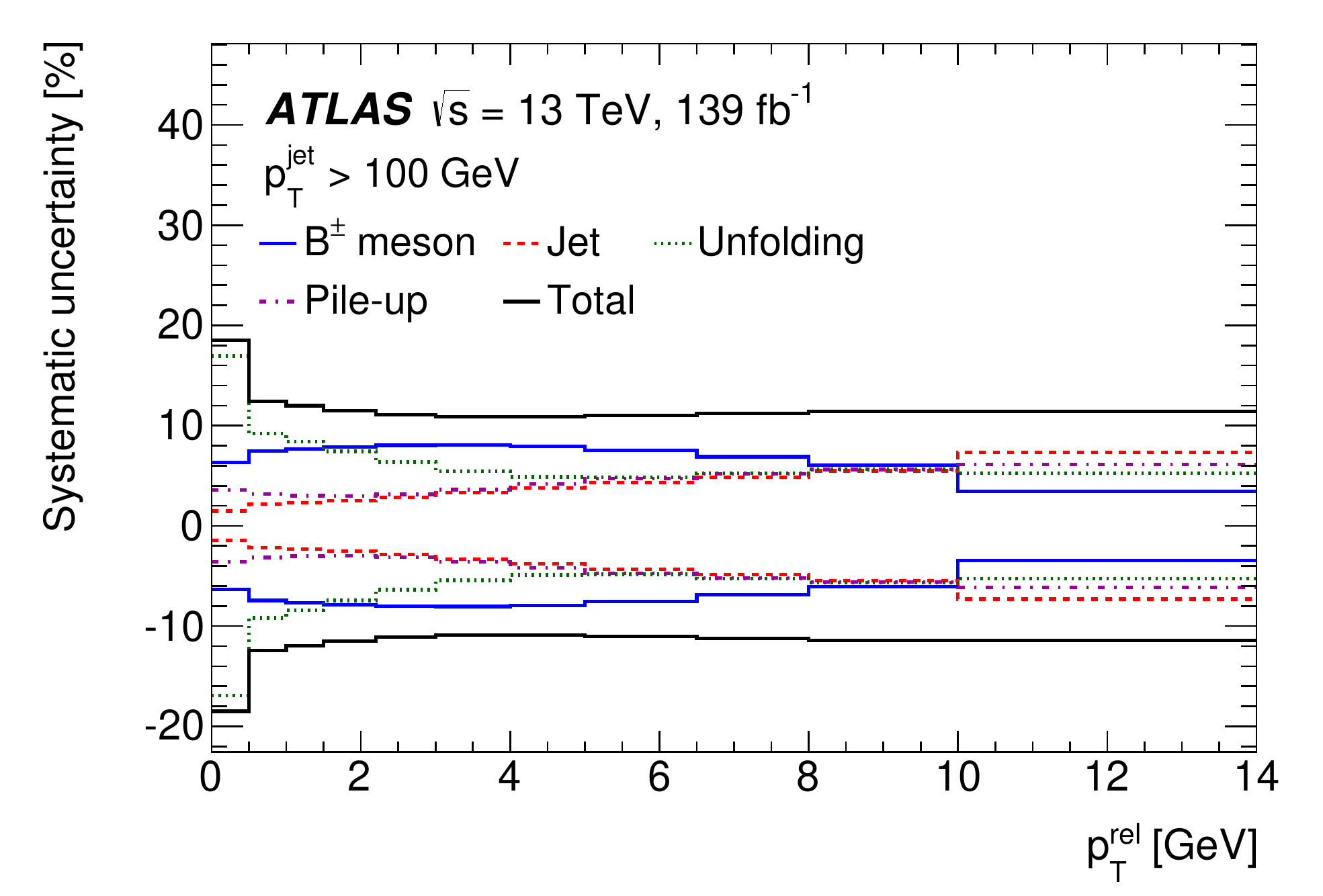}
\caption{Systematic uncertainties as a function of $z$ for all jet-$\pt$ bins (left) and as a function of $\ptrel$ for all jet-$\pt$ bins (right).}
\label{fig:systematics}
\end{figure}
% End of text imported from the .//systematics.tex input file
\clearpage
\section{Results}
\label{sec:results}
% The next lines are included from the .//results.tex input file
The particle-level results are presented and compared with the predictions described in Section~\ref{sec:montecarlo}. Figures~\ref{fig:result0} to~\ref{fig:result2} show the distributions of the longitudinal and transverse profiles for each $\pt$ bin.\\
\newline
The results show important differences between the low and high $\pt$ bins. In particular, the lower tails of the $z$ distributions contain a larger fraction of the data at high $\pt$, which translates into more high-$\ptrel$ events at high $\pt$. This is understood to be due to the fact that gluon splittings, $g\to b\bar{b}$, occur with a larger probability at high values of the jet $\pt$. The resulting $b$-quarks are reconstructed in the same jet, while the $B^\pm$ meson originates from the fragmentation of one of them, thus leading to smaller values of $z$ and higher values of $\ptrel$.\\
\newline
The results for the longitudinal profile show reasonable agreement with the \HERWIG[7] prediction with the angle-ordered parton shower, while large discrepancies are observed with the dipole parton shower. In particular, the prediction largely overestimates the data in the low $z$ tails at low $\pt$, with smaller discrepancies for higher values of the jet $\pt$. This is understood to be a result of the larger fraction of jets arising from gluon splittings in the \HERWIG[7] sample with the dipole parton shower.\\
\newline
The \SHERPA predictions give a reasonable description of the $z$ distributions in the low and medium \pT bins, although both predictions differ from data for very high values of $z$, particularly for the cluster model. At high $\pt$, the description worsens for the cluster model, which tends to underestimate the data at low $z$ and significantly overestimate the data at high $z$, while the Lund string model confirms the deviation from data at high $z$.\\
\newline
The \PYTHIA[8] samples tend to give descriptions that are qualitatively similar, and these are compatible with data within the systematic uncertainties across the different jet $\pt$ bins. However, the \Monash\!+Peterson model overestimates the data at intermediate $z$ and low $\pt$, while it tends to conform with the rest at high $\pt$. The \Monash\!+Lund--Bowler and \textsc{A14}+Lund--Bowler models both overestimate the data at very high values of $z$ in all $\pt$ ranges.\\
\newline
The discrepancies observed for the $z$ distributions have their counterparts in the $\ptrel$ distributions as follows. Due to the larger gluon splitting fractions, the \HERWIG[7] sample with the dipole parton shower significantly overestimates the data for $\ptrel$ values between 1.5 and 4.0 GeV at low $\pt$, while the differences are smaller with increasing $\pt$. The \HERWIG angle-ordered parton shower gives a better description of the $\ptrel$ distributions, although non-negligible discrepancies are also observed.\\
\newline
The \SHERPA predictions, particularly the one from the sample interfaced with the cluster hadronisation model, show large discrepancies for low values of $\ptrel$, which increase when moving towards higher bins of jet $\pt$. The description of the $\ptrel$ tails is within uncertainties at low $\pt$, but for higher $\pt$, both \SHERPA samples tend to underestimate the data for high $\ptrel$. In general, the description from the sample interfaced to the Lund string model is observed to be better than the one provided by the cluster model.\\
\newline
All \PYTHIA[8] samples give a good description of the $\ptrel$ distributions across the different jet-$\pt$ bins.
\begin{figure}[H]
\includegraphics[width=0.50\textwidth]{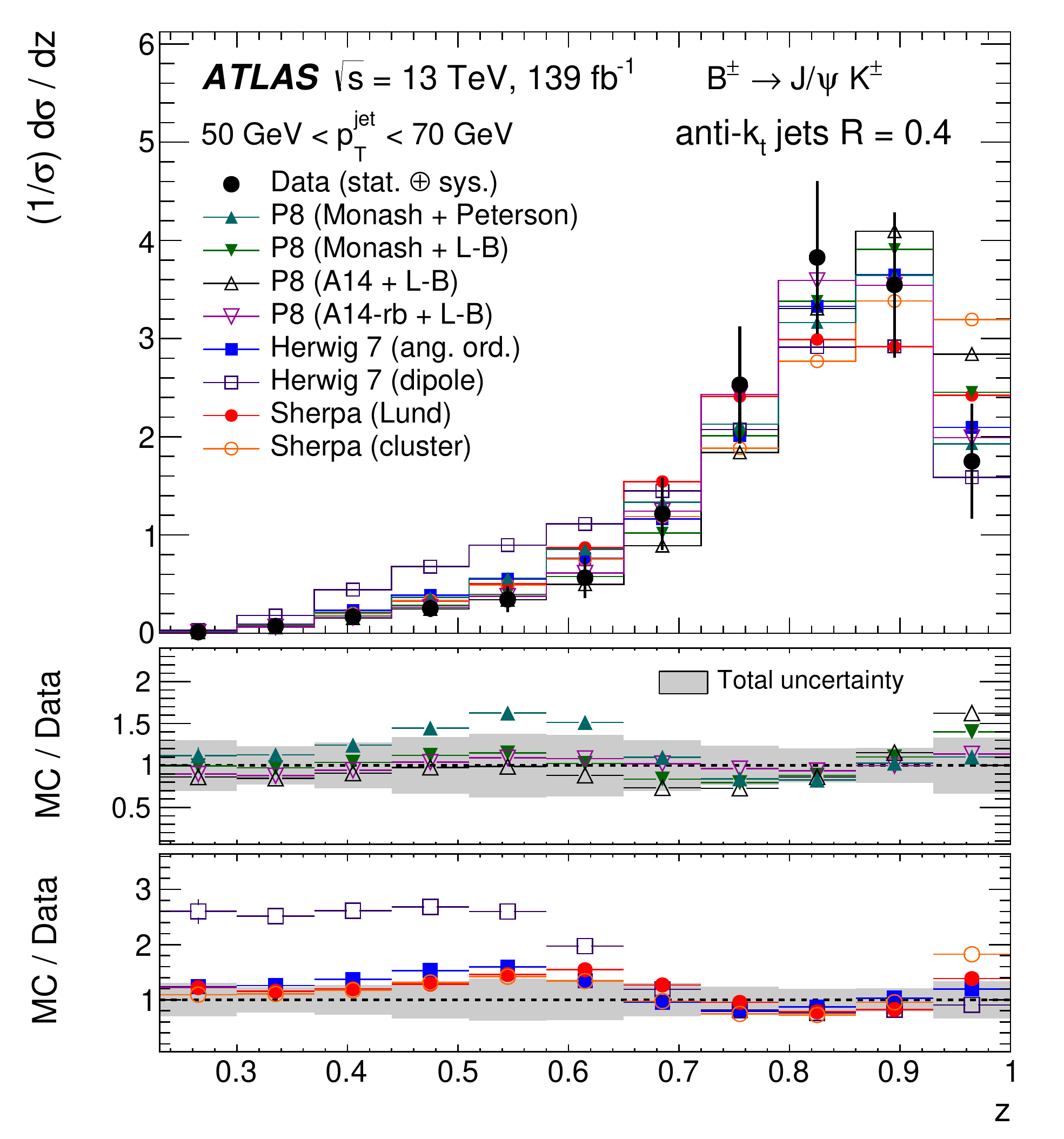}
\includegraphics[width=0.50\textwidth]{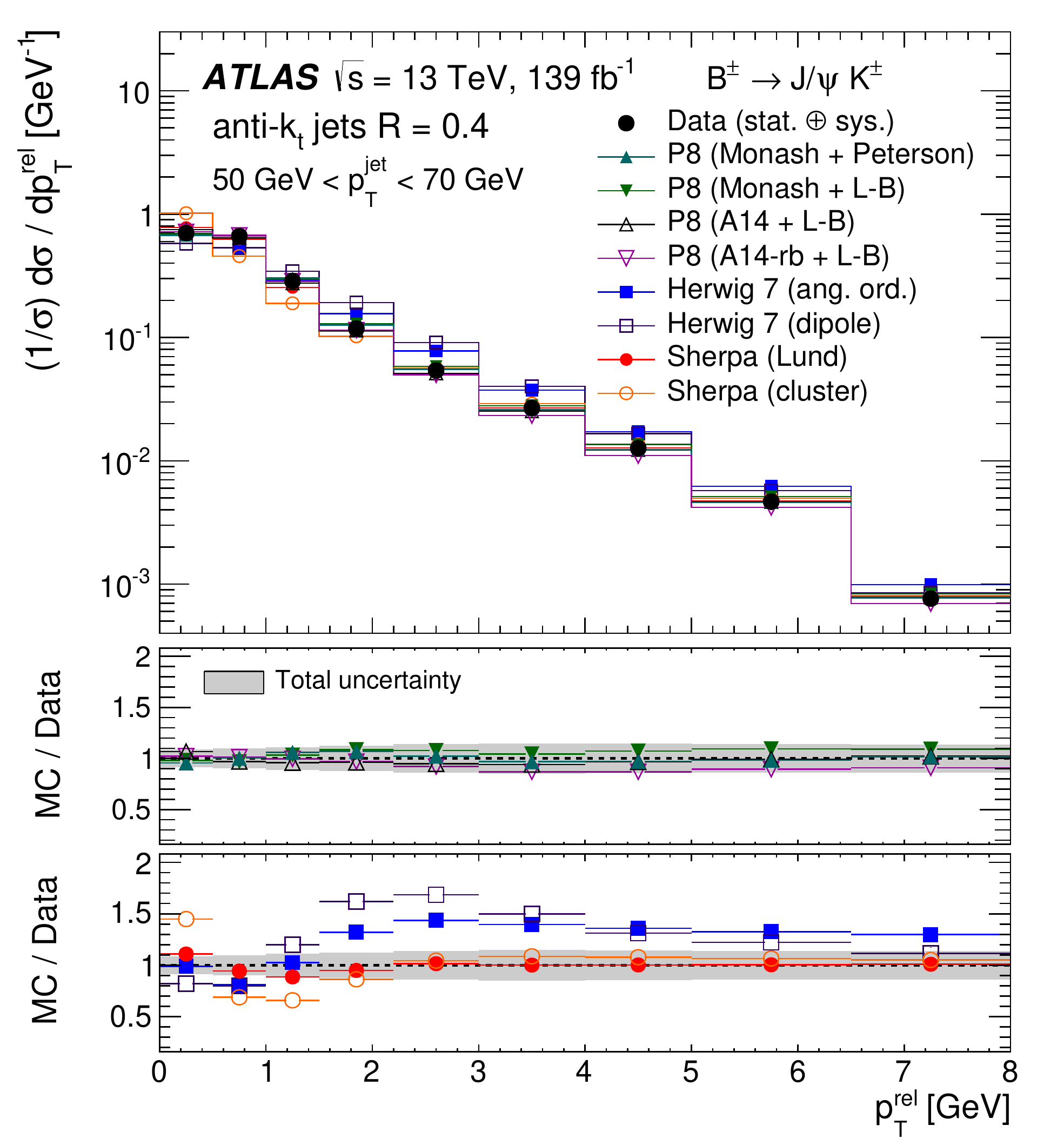}
\caption{Distributions of the longitudinal profile $z$ and the transverse profile $\ptrel$ for $50~\GeV < \pt < 70~\GeV$, together with different predictions from \PYTHIA[8], \SHERPA and \HERWIG[7]. The vertical error bars represent the total experimental uncertainties. The lower panels show the ratios of MC predictions to the data, where the gray bands represent the total uncertainties.}
\label{fig:result0}
\end{figure}
 
\begin{figure}[H]
\includegraphics[width=0.50\textwidth]{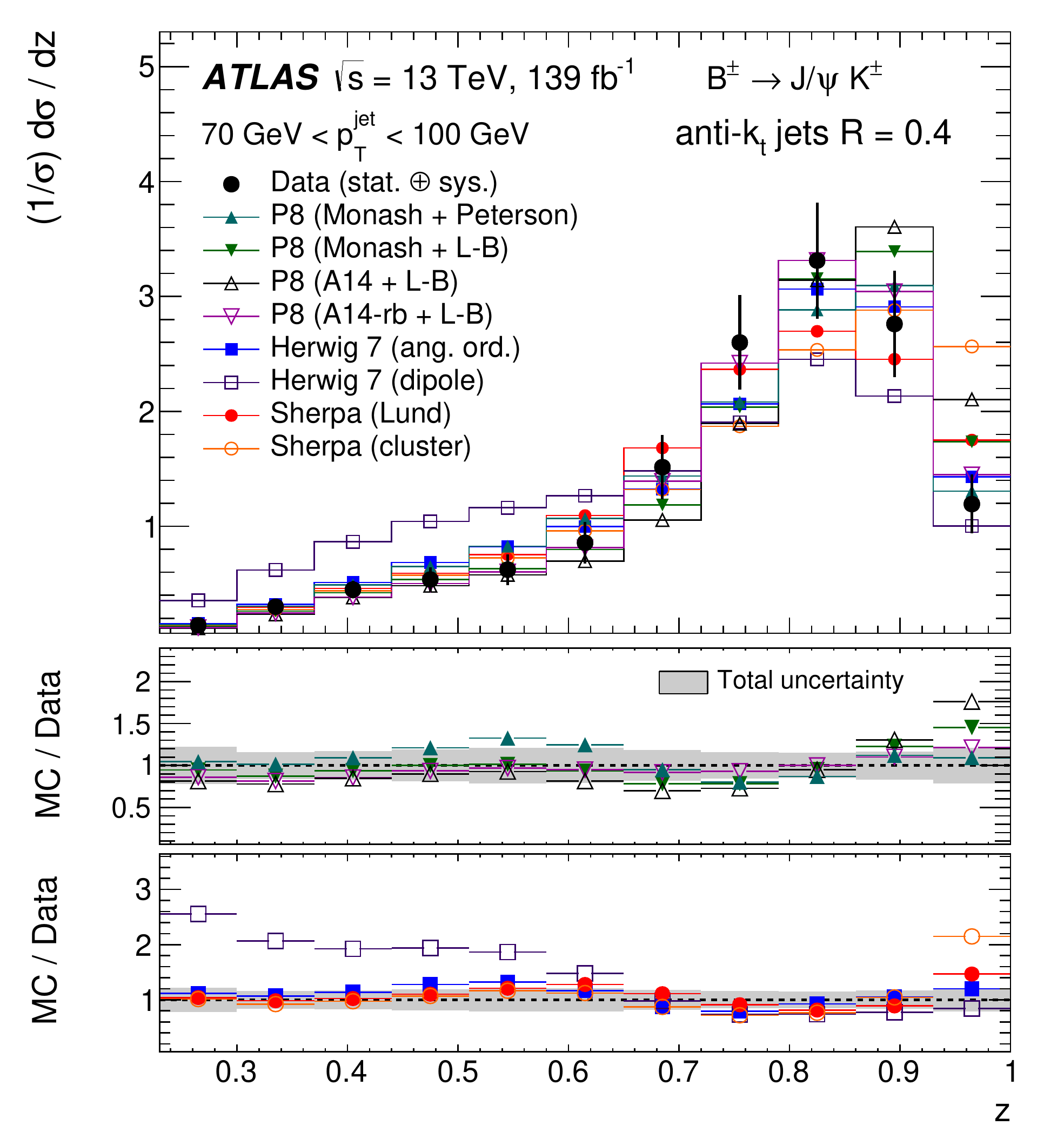}
\includegraphics[width=0.50\textwidth]{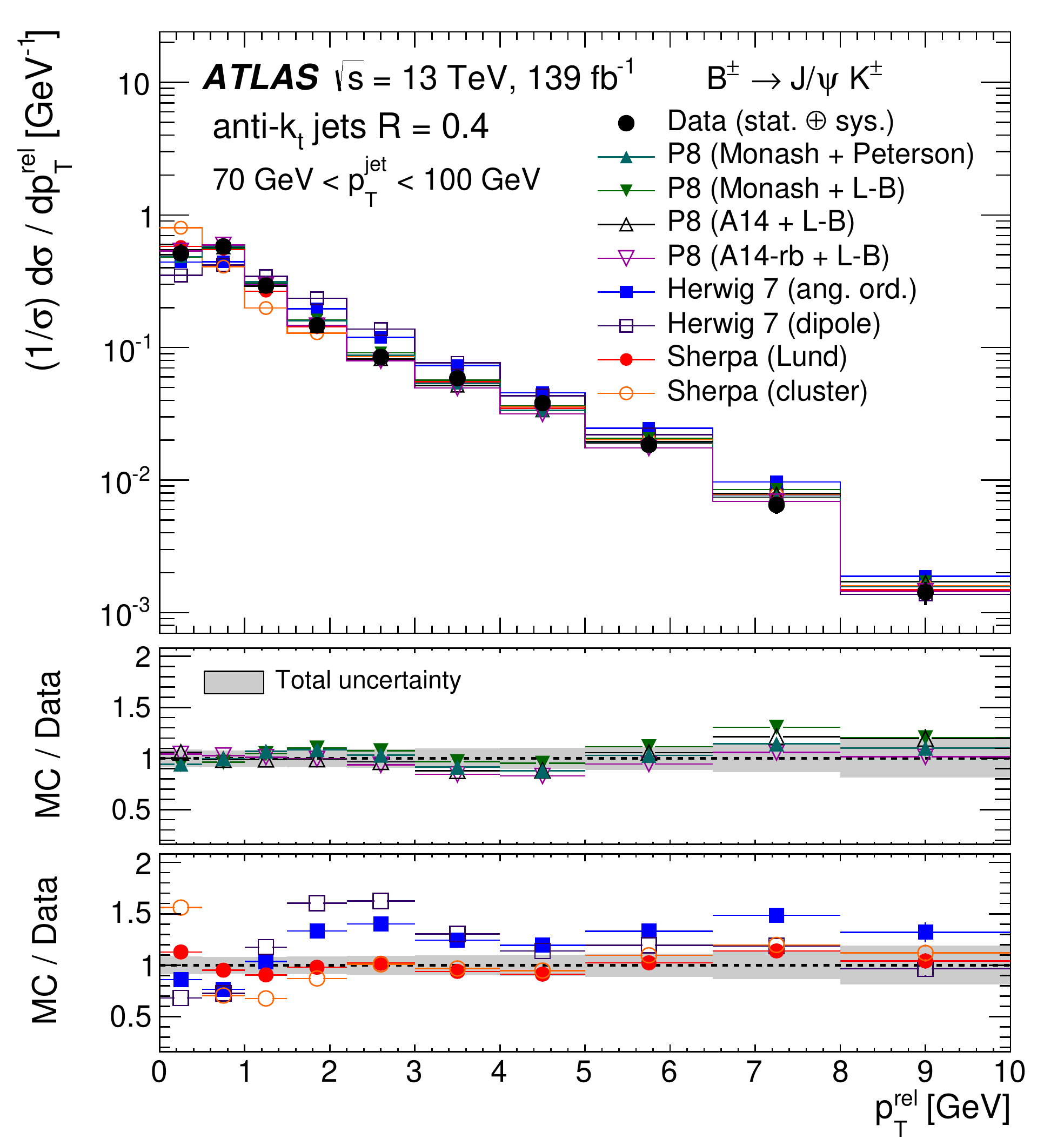}
\caption{Distributions of the longitudinal profile $z$ and the transverse profile $\ptrel$ for $70~\GeV < \pt < 100~\GeV$, together with different predictions from \PYTHIA[8], \SHERPA and \HERWIG[7]. The vertical error bars represent the total experimental uncertainties. The lower panels show the ratios of MC predictions to the data, where the gray bands represent the total uncertainties.}
\label{fig:result1}
\end{figure}
 
\begin{figure}[H]
\vspace{0.4cm}
\includegraphics[width=0.50\textwidth]{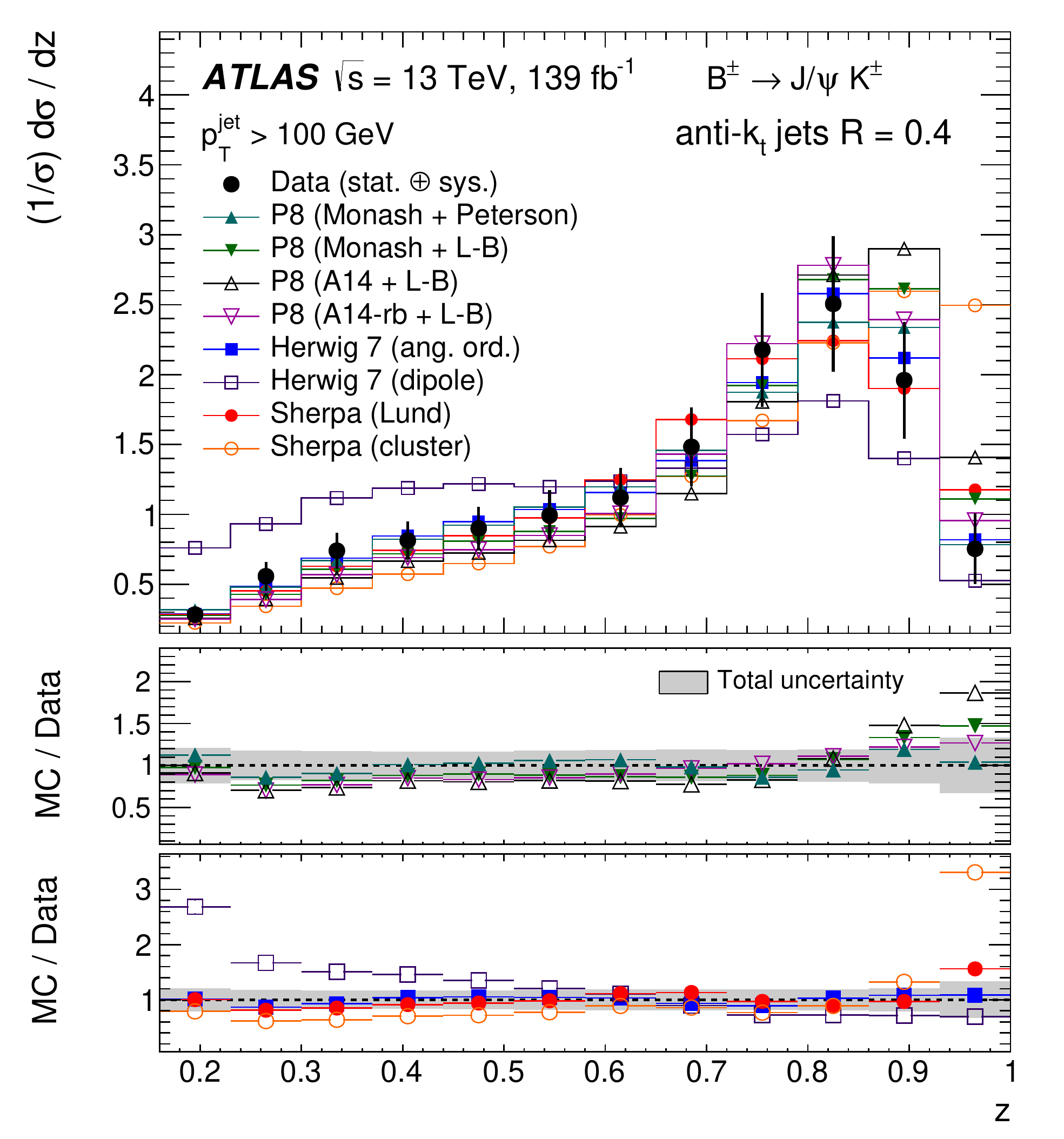}
\includegraphics[width=0.50\textwidth]{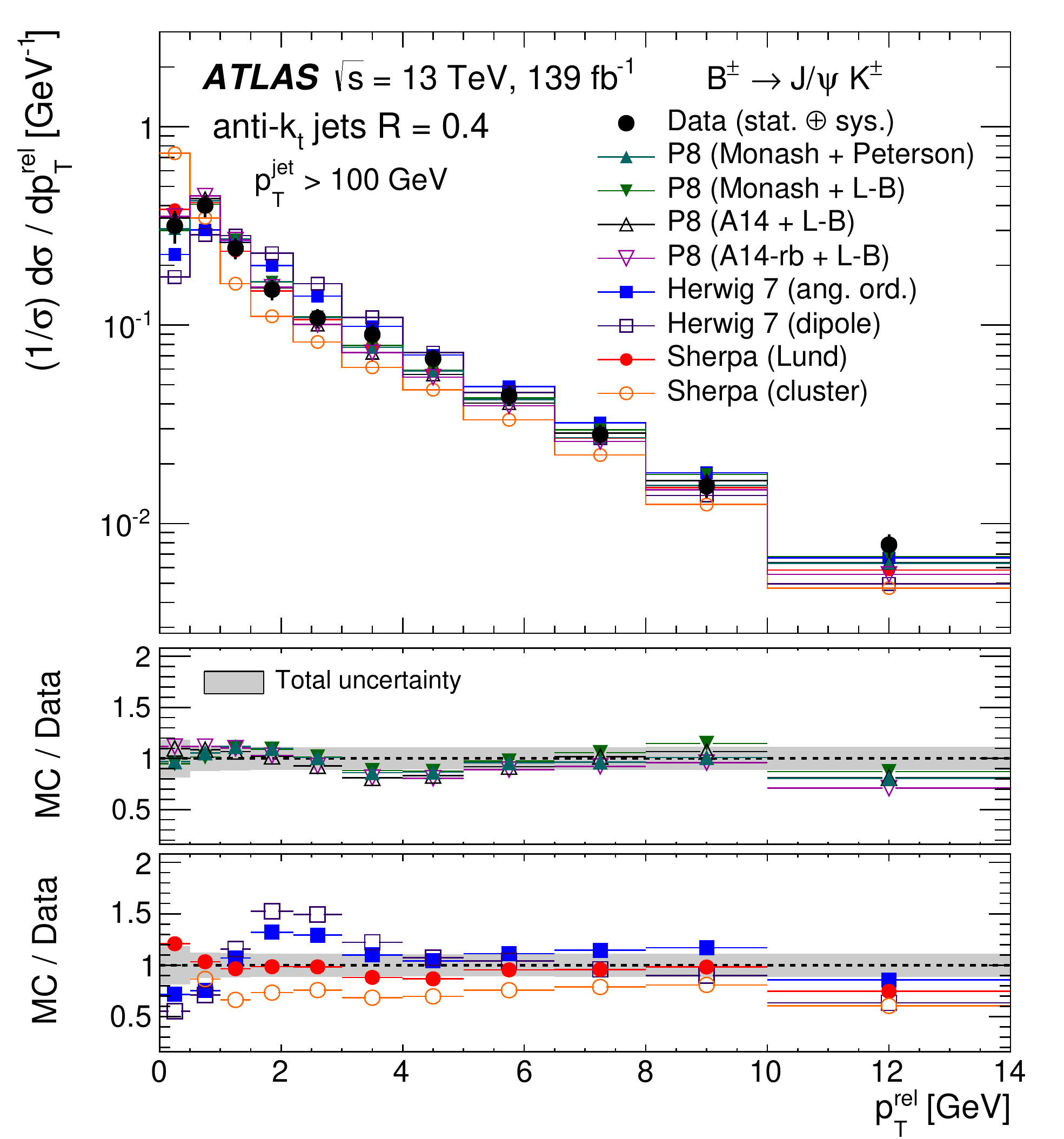}
\caption{Distributions of the longitudinal profile $z$ and the transverse profile $\ptrel$ for $\pt > 100~\GeV$, together with different predictions from \PYTHIA[8], \SHERPA and \HERWIG[7]. The vertical error bars represent the total experimental uncertainties. In the first bin of the $\ptrel$ distribution, the prediction by the \SHERPA sample using the cluster model is outside the range of the ratio panel. The lower panels show the ratios of MC predictions to the data, where the gray bands represent the total uncertainties.}
\label{fig:result2}
\end{figure}

In order to explicitly test the scale dependence of the longitudinal and transverse profiles, the average values of the longitudinal and transverse profiles, $\langle z\rangle$ and $\langle\ptrel\rangle$, are studied as a function of the jet $\pt$. The results, together with the MC predictions, are shown in Figure~\ref{fig:avgFrag}. All \PYTHIA samples describe the scale dependence reasonably well for both $\langle z\rangle$ and $\langle\ptrel\rangle$, although the samples using the \textsc{A14} and \textsc{A14-rb} tunes predict slightly larger values of $\langle z\rangle$ (and slightly lower values of $\langle\ptrel\rangle$) than measured in data. The \SHERPA sample making use of the cluster hadronisation model fails to describe the $\langle\ptrel\rangle$ data, disagreeing by 10\% to 25\%, but describes the $\langle z\rangle$ distribution reasonably well, except at high $\pt$. The \SHERPA sample interfaced to the Lund string hadronisation model describes the $\langle z\rangle$ data well, while showing small discrepancies for $\langle\ptrel\rangle$, although much smaller than when using the cluster model. The \HERWIG[7] sample making use of the angle-ordered parton shower describes the $\langle z\rangle$ scale dependence very well, while showing discrepancies of up to 15\% for the $\langle\ptrel\rangle$ data. Finally, the \HERWIG[7] sample implementing the dipole-based parton shower fails to describe both the $\langle z\rangle$ and $\langle\ptrel\rangle$ profiles, showing discrepancies of up to 10\% for the former, and up to 20\% for the latter.
 
\begin{figure}[H]
\centering
\includegraphics[width=0.49\textwidth]{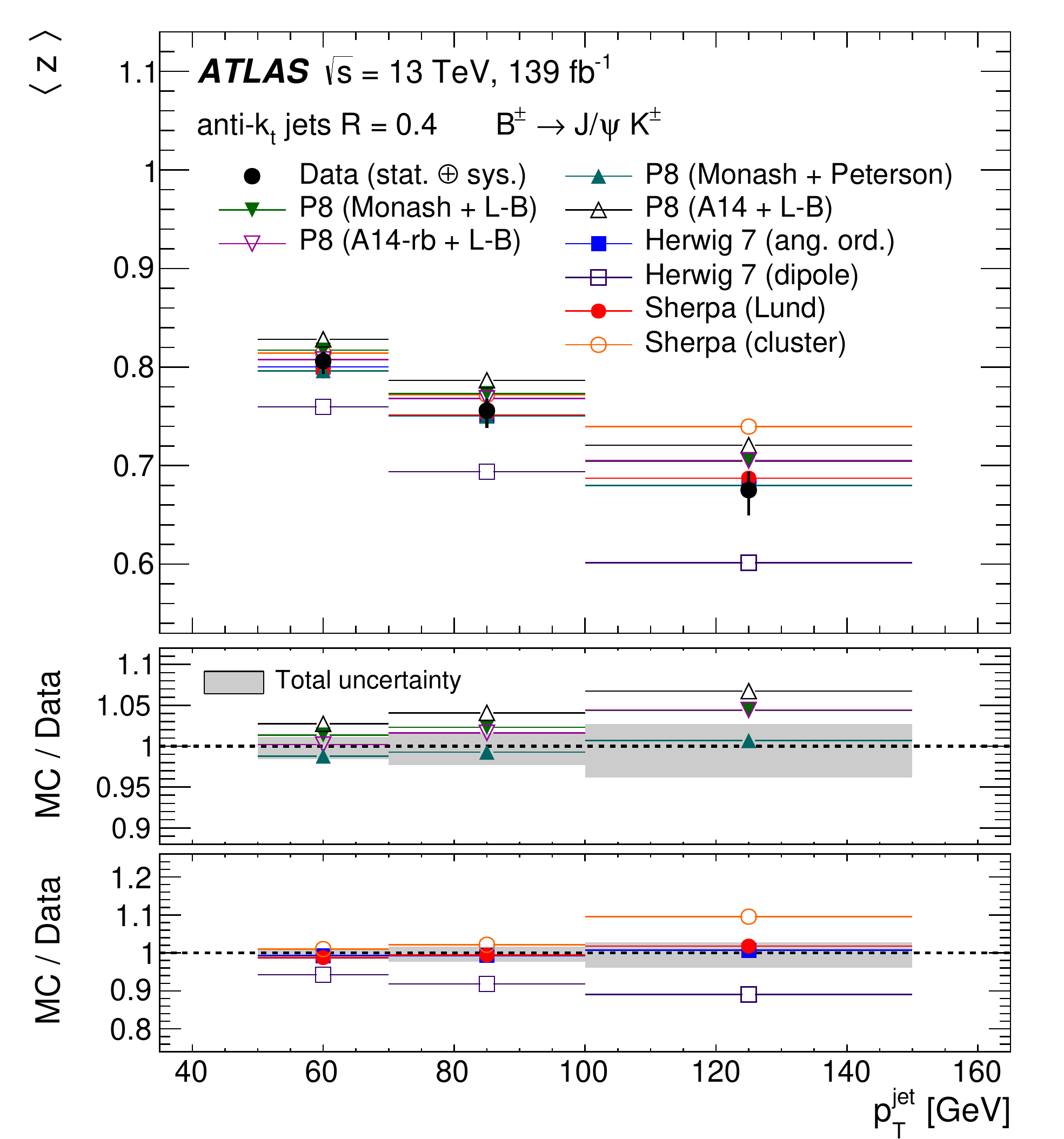}
\includegraphics[width=0.49\textwidth]{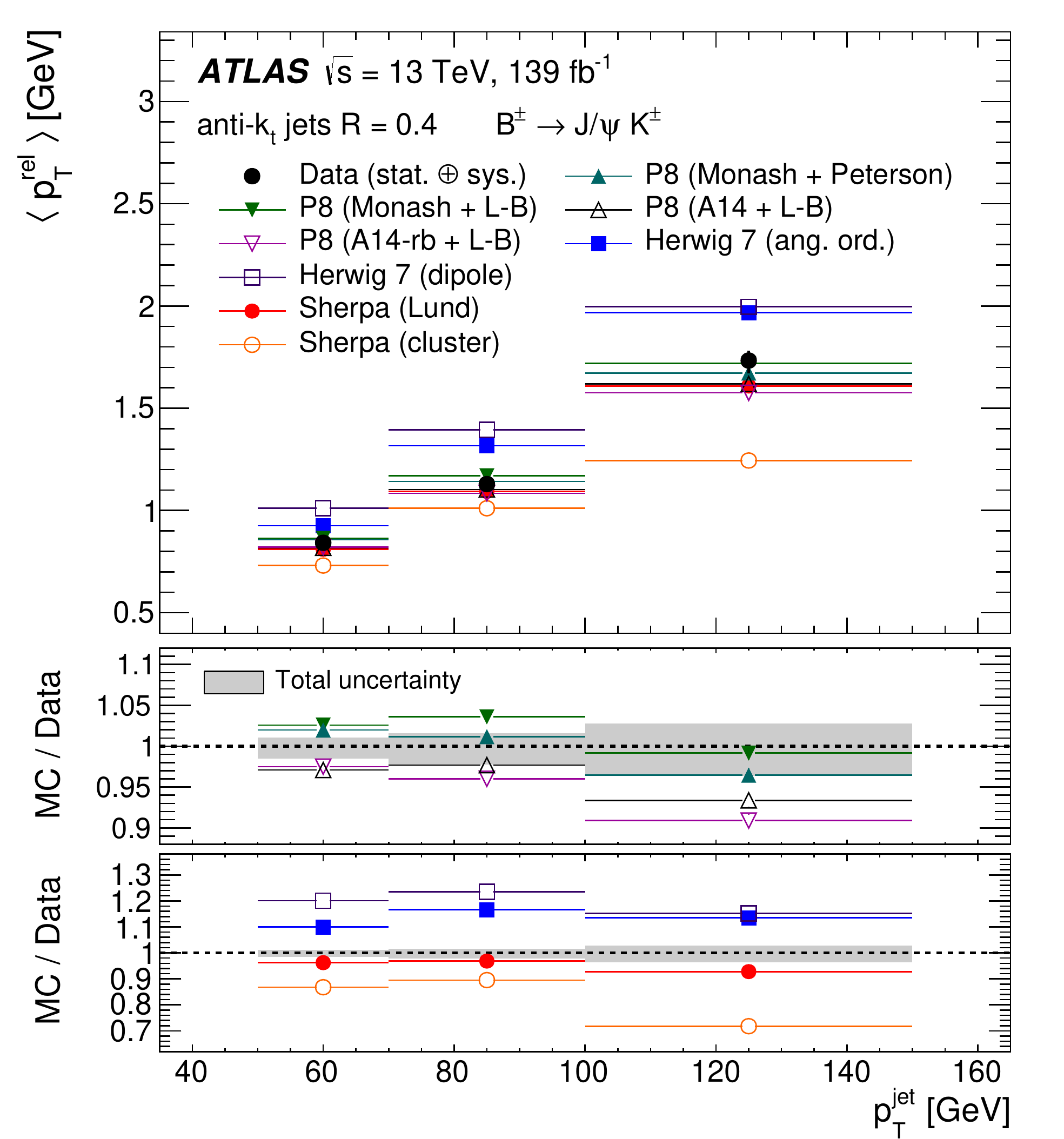}
\caption{Average values of the longitudinal profile $\langle z\rangle$ and of the transverse profile $\langle\ptrel\rangle$ as a function of the jet $\pt$, compared with MC predictions by \PYTHIA[8], \SHERPA and \HERWIG[7]. The points are in the bin centre. The vertical error bars represent the total experimental uncertainties. The lower panels show the ratios of MC predictions to the data, where the gray bands represent the total uncertainties.}
\label{fig:avgFrag}
\end{figure}
 
Some of the discrepancies described above can be traced to the modelling of the gluon splittings $g\to b\bar{b}$ in the parton shower algorithms. Jets in the MC simulation are considered to contain a gluon splitting into a $b\bar{b}$ pair if they contain two weakly decaying $B$ hadrons within $\Delta R=0.4$ of the jet axis. Figure~\ref{fig:gbbFraction} shows the fraction of such jets, for each of the MC simulations investigated, as a function of the jet $\pt$.\\
\newline
As expected, the incidence of gluon splittings in the MC simulation grows as the jet $\pt$ increases. The \PYTHIA samples show similar fractions of $g\to b\bar{b}$ jets, although the samples making use of the \Monash tune present a slightly higher fraction due to using a larger value of \alphas than the others and, hence, having a larger amount of gluon radiation. Both \SHERPA samples present larger fractions than the \PYTHIA samples for low and medium $\pt$. However, for the highest $\pt$ bin, the \SHERPA sample using the cluster model presents a lower fraction than \PYTHIA. The angle-ordered shower implemented in \HERWIG[7] shows an amount of $g\to b\bar{b}$ splitting very similar to that of \SHERPA, while the \HERWIG[7] sample making use of the dipole-based parton shower presents a much larger fraction of jets arising from gluon splittings. As pointed out previously, these differences help to explain the discrepancies between this sample and the data, since the lower tails of the $z$ distributions, as well as the higher tails for $\ptrel$, are strongly influenced by the fraction of gluon splittings.
 
\begin{figure}[H]
\centering
\includegraphics[width=0.85\textwidth]{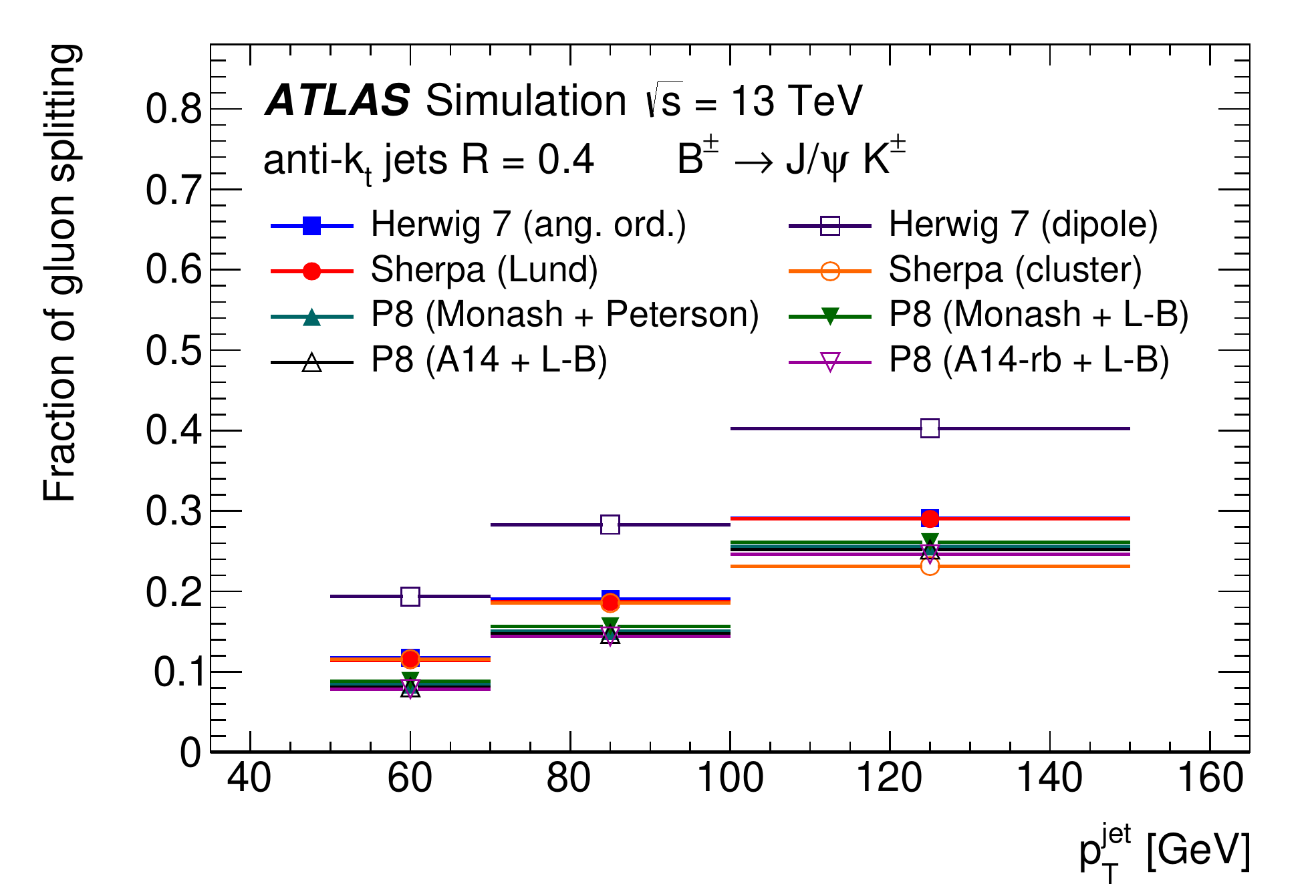}
\caption{Fraction of jets arising from gluon splittings $g\to b\bar{b}$ as a function of the jet $\pt$ for different MC models.}
\label{fig:gbbFraction}
\end{figure}
% End of text imported from the .//results.tex input file
\section{Summary and conclusions}
\label{sec:summary}
% The next lines are included from the .//summary.tex input file
The fragmentation properties of $b$-jets have been measured by using the decay of charged $B$ mesons into $J/\psi$ charmonia and charged kaons, using 139~\ifb\ of $pp$ collisions at $\sqrt{s} = 13~\TeV$ recorded by the ATLAS detector at the LHC. To this end, the decay channel of the $B$ mesons is fully reconstructed using a three-track fit to a common vertex. The $B^\pm$ candidates are matched to jets, reconstructed using the anti-$k_t$ algorithm with $R=0.4$, and the longitudinal and transverse profiles of the $B$ mesons over the jet momentum are measured. The measurement spans three bins of the jet transverse momentum, and the yield of $B$ mesons is extracted for each bin of the fragmentation variable ($z$ or $\ptrel$) and the jet $\pt$. The results are then corrected for detector resolution effects using an iterative Bayesian algorithm, and the systematic uncertainties due to the muon, jet and event reconstruction properties are evaluated.\\
\newline
The results are compared with different MC predictions, including \PYTHIA[8], \HERWIG[7] and \SHERPA samples using different models for the parton shower and hadronisation. Generally, the best description of the longitudinal profile is provided by the \PYTHIA[8] and \SHERPA samples making use of the string hadronisation model, which provide similar descriptions for all values of the jet $\pt$. For medium and high values of the jet $\pt$, the \SHERPA sample with cluster hadronisation shows large deviations from data at very high values of $z$. The \HERWIG[7] sample with the angle-ordered parton shower also gives a good description of the longitudinal profile for all values of the jet $\pt$. In contrast, the \HERWIG[7] samples using the dipole-based parton shower show large deviations from the data in the lower tails of the $z$ distribution. For the transverse profile, similar comments apply. While the \PYTHIA[8] and \SHERPA samples using the string hadronisation model give fair and similar descriptions of the data for all values of the jet $\pt$, \HERWIG[7] fails to describe the data. The angle-ordered parton shower, however, provides a better description than the dipole-based parton shower. At medium and high $\pt$, the \SHERPA sample with the cluster hadronisation model gives a poor description for low values of $\ptrel$.
\newline
This analysis provides key measurements which help to better understand the fragmentation functions of heavy quarks. As has been shown, significant differences among different MC models are observed, and also between the models and the data. Some of the discrepancies are understood to arise from poor modelling of the $g\to b\bar{b}$ splittings, to which the present analysis has substantial sensitivity. Including the present measurements in a future tune of the MC predictions may help to improve the description and reduce the theoretical uncertainties of processes where heavy-flavour quarks are present in the final state, such as top quark pair production or Higgs boson decays into heavy quark pairs.
% End of text imported from the .//summary.tex input file
 
\section*{Acknowledgements}
% The next lines are included from the .//acknowledgements/Acknowledgements.tex input file

We thank CERN for the very successful operation of the LHC, as well as the
support staff from our institutions without whom ATLAS could not be
operated efficiently.
 
We acknowledge the support of
ANPCyT, Argentina;
YerPhI, Armenia;
ARC, Australia;
BMWFW and FWF, Austria;
ANAS, Azerbaijan;
SSTC, Belarus;
CNPq and FAPESP, Brazil;
NSERC, NRC and CFI, Canada;
CERN;
ANID, Chile;
CAS, MOST and NSFC, China;
Minciencias, Colombia;
MEYS CR, Czech Republic;
DNRF and DNSRC, Denmark;
IN2P3-CNRS and CEA-DRF/IRFU, France;
SRNSFG, Georgia;
BMBF, HGF and MPG, Germany;
GSRI, Greece;
RGC and Hong Kong SAR, China;
ISF and Benoziyo Center, Israel;
INFN, Italy;
MEXT and JSPS, Japan;
CNRST, Morocco;
NWO, Netherlands;
RCN, Norway;
MEiN, Poland;
FCT, Portugal;
MNE/IFA, Romania;
JINR;
MES of Russia and NRC KI, Russian Federation;
MESTD, Serbia;
MSSR, Slovakia;
ARRS and MIZ\v{S}, Slovenia;
DSI/NRF, South Africa;
MICINN, Spain;
SRC and Wallenberg Foundation, Sweden;
SERI, SNSF and Cantons of Bern and Geneva, Switzerland;
MOST, Taiwan;
TAEK, Turkey;
STFC, United Kingdom;
DOE and NSF, United States of America.
In addition, individual groups and members have received support from
BCKDF, CANARIE, Compute Canada and CRC, Canada;
COST, ERC, ERDF, Horizon 2020 and Marie Sk{\l}odowska-Curie Actions, European Union;
Investissements d'Avenir Labex, Investissements d'Avenir Idex and ANR, France;
DFG and AvH Foundation, Germany;
Herakleitos, Thales and Aristeia programmes co-financed by EU-ESF and the Greek NSRF, Greece;
BSF-NSF and GIF, Israel;
Norwegian Financial Mechanism 2014-2021, Norway;
NCN and NAWA, Poland;
La Caixa Banking Foundation, CERCA Programme Generalitat de Catalunya and PROMETEO and GenT Programmes Generalitat Valenciana, Spain;
G\"{o}ran Gustafssons Stiftelse, Sweden;
The Royal Society and Leverhulme Trust, United Kingdom.
 
The crucial computing support from all WLCG partners is acknowledged gratefully, in particular from CERN, the ATLAS Tier-1 facilities at TRIUMF (Canada), NDGF (Denmark, Norway, Sweden), CC-IN2P3 (France), KIT/GridKA (Germany), INFN-CNAF (Italy), NL-T1 (Netherlands), PIC (Spain), ASGC (Taiwan), RAL (UK) and BNL (USA), the Tier-2 facilities worldwide and large non-WLCG resource providers. Major contributors of computing resources are listed in Ref.~\cite{ATL-SOFT-PUB-2021-003}.
 
% End of text imported from the .//acknowledgements/Acknowledgements.tex input file
 
\clearpage
 
\printbibliography

\clearpage
 
% ATLAS Collaboration author list
% Reference date of STDM-2018-52 is 2021-01-26
% Author list last updated on date 16-NOV-21
% Data extracted on 16-Nov-2021 for paper reference STDM-2018-52
% at 4:50pm
 
\begin{flushleft}
\hypersetup{urlcolor=black}
{\Large The ATLAS Collaboration}

\bigskip

\AtlasOrcid[0000-0002-6665-4934]{G.~Aad}$^\textrm{\scriptsize 99}$,    
\AtlasOrcid[0000-0002-5888-2734]{B.~Abbott}$^\textrm{\scriptsize 126}$,    
\AtlasOrcid[0000-0002-7248-3203]{D.C.~Abbott}$^\textrm{\scriptsize 100}$,    
\AtlasOrcid[0000-0002-2788-3822]{A.~Abed~Abud}$^\textrm{\scriptsize 34}$,    
\AtlasOrcid[0000-0002-1002-1652]{K.~Abeling}$^\textrm{\scriptsize 51}$,    
\AtlasOrcid[0000-0002-2987-4006]{D.K.~Abhayasinghe}$^\textrm{\scriptsize 91}$,    
\AtlasOrcid[0000-0002-8496-9294]{S.H.~Abidi}$^\textrm{\scriptsize 27}$,    
\AtlasOrcid[0000-0002-8279-9324]{O.S.~AbouZeid}$^\textrm{\scriptsize 38}$,    
\AtlasOrcid[0000-0001-5329-6640]{H.~Abramowicz}$^\textrm{\scriptsize 159}$,    
\AtlasOrcid[0000-0002-1599-2896]{H.~Abreu}$^\textrm{\scriptsize 158}$
% ...
\bigskip
\\

% ...
$^{27}$Physics Department, Brookhaven National Laboratory, Upton NY; United States of America.\\
% ...
$^{34}$CERN, Geneva; Switzerland.\\
% ...
$^{38}$Niels Bohr Institute, University of Copenhagen, Copenhagen; Denmark.\\
% ...
$^{51}$II. Physikalisches Institut, Georg-August-Universit\"{a}t G\"ottingen, G\"ottingen; Germany.\\
% ...
$^{91}$Department of Physics, Royal Holloway University of London, Egham; United Kingdom.\\
% ...
$^{99}$CPPM, Aix-Marseille Universit\'e, CNRS/IN2P3, Marseille; France.\\
$^{100}$Department of Physics, University of Massachusetts, Amherst MA; United States of America.\\
% ...
$^{126}$Homer L. Dodge Department of Physics and Astronomy, University of Oklahoma, Norman OK; United States of America.\\
% ...
$^{158}$Department of Physics, Technion, Israel Institute of Technology, Haifa; Israel.\\
$^{159}$Raymond and Beverly Sackler School of Physics and Astronomy, Tel Aviv University, Tel Aviv; Israel.\\
% ...

\end{flushleft}

% Created with Glance <Atlas.Glance@cern.ch>

\end{document}